\documentclass[a4paper,11pt]{article}
\pdfoutput=1 

\usepackage{jheppub} 
\usepackage{graphicx,color}
\usepackage{amsmath}
\usepackage{slashed}
\usepackage{multirow}
\usepackage{multicol,listings}
\usepackage{autobreak}
\usepackage[normalem]{ulem}
\usepackage{hyperref}
\usepackage{cleveref}
\usepackage{slashed}
\usepackage{amssymb}
\usepackage{epsfig}
\usepackage{float}

\usepackage[utf8]{inputenc}
    \usepackage[T1]{fontenc}
    \usepackage{lmodern}
    \usepackage[x11names]{xcolor}
    \usepackage{framed}

\colorlet{shadecolor}{blue!10}

\SetSymbolFont{largesymbols}{bold}{OMX}{txex}{b}{n}

\DeclareMathAlphabet{\mathpzc}{OT1}{pzc}{m}{it}

 \definecolor{jd}{rgb}{0.858, 0.188, 0.478}

\usepackage{bbm}  
\usepackage{amsmath}                                                

\newcommand{\nc}{\newcommand}
\nc{\beq}{\begin{equation}}  \nc{\eeq}{\end{equation}}
\nc{\bea}{\begin{eqnarray}}  \nc{\eea}{\end{eqnarray}}
\nc{\baa}{\begin{array}}     \nc{\eaa}{\end{array}}
\nc{\bit}{\begin{itemize}}   \nc{\eit}{\end{itemize}}
\nc{\ben}{\begin{enumerate}} \nc{\een}{\end{enumerate}}
\nc{\bce}{\begin{center}}    \nc{\ece}{\end{center}}
\nc{\bpm}{\begin{pmatrix}}   \nc{\epm}{\end{pmatrix}}
\nc{\bvt}{\begin{verbatim}}  \nc{\evt}{\end{verbatim}}
\nc{\bal}{\begin{align}}
\def\mcr{\nonumber\\[6pt]}

\def\to{\rightarrow}

\def\boldoverdot{\,{\raise6pt\hbox{\bf.}\!\!\!\!\>}}
\def\re{{\bf Re}}
\def\im{{\bf Im}}

\def\then{{\quad\Rightarrow\quad}}

\def\lcal{{\cal L}}

\def\ocal{{\cal O}}

\usepackage{bm}
\def\diag{\hbox{\diag}}

\def\gev{\hbox{GeV}}
\def\tev{\hbox{TeV}}

\def\doubleundertext#1{
{\undertext{\vphantom{y}#1}}\par\nobreak\vskip-\the\baselineskip\vskip4pt%
\undertext{\hbox to 2in{}}}
\def\inbox#1{\vbox{\hrule\hbox{\vrule\kern5pt
     \vbox{\kern5pt#1\kern5pt}\kern5pt\vrule}\hrule}}
\def\sqr#1#2{{\vcenter{\hrule height.#2pt
      \hbox{\vrule width.#2pt height#1pt \kern#1pt
         \vrule width.#2pt}
      \hrule height.#2pt}}}

\def\today{\ifcase\month\or
  January\or February\or March\or April\or May\or June\or
  July\or August\or September\or October\or November\or December\fi
  \space\number\day, \number\year}
\def\pmb#1{\setbox0=\hbox{#1}%
  \kern-.025em\copy0\kern-\wd0
  \kern.05em\copy0\kern-\wd0
  \kern-.025em\raise.0433em\box0 }

\def\pmbb#1{\setbox0=\hbox{#1}%
  \kern-.02em\copy0\kern-\wd0
  \kern.04em\copy0\kern-\wd0
  \kern-.02em\raise.03464em\box0 }
\def\up#1{^{\left( #1 \right) }}
\def\inv#1{\frac1{#1}}
\def\su#1{{SU(#1)}}

\def\sumprime_#1{\setbox0=\hbox{$\scriptstyle{#1}$}
  \setbox2=\hbox{$\displaystyle{\sum}$}
  \setbox4=\hbox{${}'\mathsurround=0pt$}
  \dimen0=.5\wd0 \advance\dimen0 by-.5\wd2
  \ifdim\dimen0>0pt
  \ifdim\dimen0>\wd4 \kern\wd4 \else\kern\dimen0\fi\fi
\mathop{{\sum}'}_{\kern-\wd4 #1}}
\def\lapp{\mathrel{\rlap{\raise.5ex\hbox{$<$}}
                    {\lower.5ex\hbox{$\sim$}}}}
\def\gapp{\mathrel{\rlap{\raise.5ex\hbox{$>$}}
                    {\lower.5ex\hbox{$\sim$}}}}

{
{

\def\ocal{{\cal O}}

\def\phit{\tilde\phi}

\def\wc{C} 
\def\smvev{{\tt v}} 

\newcommand{\bmt}{\begin{pmatrix}}
\newcommand{\emt}{\end{pmatrix}}

\newcommand{\baz}{\begin{array}{cc}}
\newcommand{\mathsym}[1]{{}}

\newcommand{\bt}{\begin{tabular}}
\newcommand{\et}{\end{tabular}}

\newcommand{\benu}{\begin{enumerate}}
\newcommand{\eenu}{\end{enumerate}}
\newcommand{\bav}{\begin{array}{cccc}}

\title{\boldmath Associated production of Higgs and single top at the LHC in presence of the SMEFT operators}

\author[a]{Subhaditya Bhattacharya,}
\author[b]{Sanjoy Biswas,}
\author[c]{Kuntal Pal,}
\author[c]{Jose Wudka}

\affiliation[a]{Department of Physics, Indian Institute of Technology Guwahati, Assam 781039, India}
\affiliation[b]{Department of Physics, Ramakrishna Mission Vivekananda Educational and Research Institute, Belur Math, Howrah 711202, India}
\affiliation[c]{ Department of Physics and Astrophysics, University of California, Riverside, USA}
\emailAdd{subhab@iitg.ac.in}
\emailAdd{sanjoy.phy@gm.rkmvu.ac.in}
\emailAdd{kpal002@ucr.edu}
\emailAdd{jose.wudka@ucr.edu}

\abstract{We analyse the single top production in association with the Higgs at the Large Hadron Collider (LHC) using Standard Model (SM) effective
operators upto dimension six. We show that the presence of effective operators can significantly alter the existing bound on the top-Higgs Yukawa 
coupling. We analyse events at the LHC with 35.9 and 137(140) fb$^{-1}$ integrated luminosities using both cut-based and machine learning techniques 
to probe new physics (NP) scale and operator coefficients addressing relevant SM background reduction. 
The four fermi effective operator(s) that contribute to the signal, turn out to be crucial and a limit on the top-Higgs Yukawa interaction in presence of them
is obtained from the present data and for future sensitivities. }


\begin{document} 
\maketitle
\flushbottom

\section{Introduction}

The discovery of the Higgs boson at the LHC completed the spectrum predicted by the Standard Model (SM). Although the 
measured properties of this $125\,\gev$  scalar strongly favour those of the SM Higgs (corresponding to a weak isodoublet 
with $Y=1/2$), we are yet to pin them down fully, with the data still allowing $\gtrsim O(10\%)$ $1$$\sigma$ differences 
(assuming Gaussian errors) from the SM predictions \cite{Workman:2022ynf}.

A statistically significant deviation of the Higgs couplings form the SM predictions would provide strong evidence of new physics (NP), which is often parameterized by 
the signal-strength coefficients $ \kappa_X $, that take the value $\kappa_X = 1$ in the SM. Accordingly, most studies 
look for evidence of NP in deviations from this value. However, if NP is present, there is no reason to believe that its only (or main) effect will be to modify the $ \kappa_X $, 
in general, the process used to measure the couplings $\kappa_X$ will also be affected by other NP effects. 

Therefore a general study of the experimental sensitivity to the $ \kappa_X$ must also include all possible NP effects relevant to the processes under consideration, 
which is often more significant than the modification of $\kappa_X$ from its SM prediction. Taking into account these other NP effects can significantly alter the limits on $ \kappa_X$; 
ignoring them leads to limits on the $ \kappa_X$ that are relevant only for very limited types of NP.

The purpose of this note is to illustrate these issues and analyze their consequences for the special case of an LHC process that is well-suited to measure the 
Higgs-top quark signal strength $ \kappa_{\tt t}$ which is not yet well constrained, $ |\kappa_{\tt t}| < 1.7 $ at 95\% C.L. (assuming no other NP effects), 
and even its sign (relative to that of the $WWh$ coupling) is unknown \cite{Workman:2022ynf}. The main channels through which the Higgs-top Yukawa coupling 
$\kappa_{\tt t}$ can be accessed are Higgs production via gluon fusion, $ht\bar{t}$ production, and $ h \to \gamma \gamma $ decay. However, apart from 
$ h \to \gamma \gamma $ decay, none of them are sensitive to the relative sign between $ht\bar{t}$ and $WWh$ couplings~{\footnote{Resolving the sign between $hZZ$ and $hWW$ 
couplings has been addressed in \cite{Xie:2021xtl} and references therein.}}. Moreover, the gluon fusion and 
the $ht\bar{t}$ production receive significant higher order QCD corrections \cite{CMS:2020cga}; while the Higgs to photon decay 
process is suppressed since it is generated only at one loop.

\begin{figure}
$$
\includegraphics[height=6.0cm]{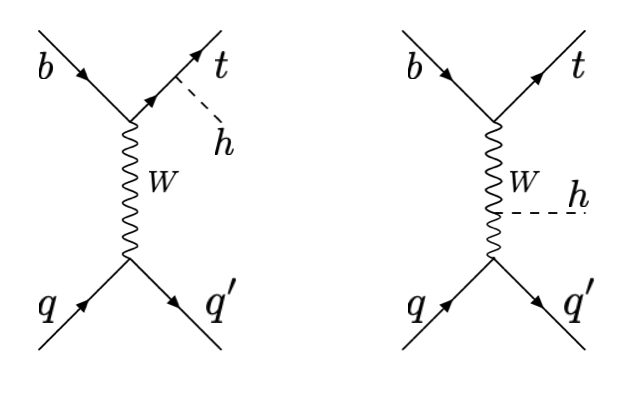}
$$
\caption{SM contributions to $t$-channel single-top production in association with a Higgs boson at the LHC.}
\label{fig:SM.thj}
\end{figure}

These problems are not shared by Higgs production in association with single-top, which is sensitive to the magnitude and sign of 
$ \kappa_{\tt t}$ \cite{Biswas:2012bd, Farina:2012xp}. In the SM there are two relevant graphs (\cref{fig:SM.thj}), one proportional to $\kappa_{\tt t} $ 
and the other to the $hWW$ coupling $g_{\tt hww}~$\footnote{For constraints on the magnitude of $g_{\tt hww} $, see \cite{Banerjee:2019twi}.}, 
so the cross-section will be particularly sensitive to the relative sign and magnitude of these two couplings. 
Moreover, the SM predicts that the two contributing diagrams will interfere destructively, making this process also sensitive to NP effects. 
In the following, we will study this process, which we refer to as $thj$ ($j$ denotes the light jet in the final state), as a probe of NP.

It was shown in \cite{Biswas:2012bd, Farina:2012xp} that, absent any NP effects, the $thj$ cross-section increases significantly when $\kappa_{\tt t}$ takes 
large negative values. The present LHC data also confirms the exclusion of $\kappa_{\tt t} < -0.9$ \cite{CMS:2018jeh}. 
The theoretical cross-section for $thj$ production in the absence of other NP effects can be parametrized in terms of $\kappa_{\tt t}$ and $\kappa_{\tt w}$ 
at NLO using {\sf MADGRAPH5@MC@NLO} \cite{Alwall:2011uj} as follows,
\beq
\sigma_{thj}=(2.63\kappa_{\tt t}^2+3.58\kappa_{\tt w}^2-5.21\kappa_{\tt t}\kappa_{\tt w})\sigma^{\rm SM}_{thj}\,, \qquad \kappa_{\tt w} = \frac{g_{\tt hww}}{g\up{\rm SM}_{\tt hww}}\,.
\eeq
$\sigma^{\rm SM}_{thj}$ at the LHC with $\sqrt{s}=13$ TeV is 71 fb after using appropriate $K$ factor of 1.187, as used by CMS analysis.
In this analysis we consider all the effective (EFT) operators upto dimension six  \cite{Grzadkowski:2010es} that contribute to the $thj$ production at LHC. Significant contribution arises 
only from those which are potentially tree generated (PTG) \cite{Einhorn:2013kja}, while the operators that are loop generated (LG) suffer additional suppression and are neglected. 
Similar studies have been done in the context of one top quark and a pair of gauge boson production in \cite{Maltoni:2019aot,Faham:2021zet}, 
for both $thj$ and $tZj$ processes at LHC including NLO corrections in \cite{Degrande:2018fog}, for finding the relative contribution of the operators that contribute to 
$thj$ in \cite{Guchait:2022ktz}. A correlation between the operators that contribute to both $pp \to tHq$ and $pp \to tq$ has been performed in \cite{Cao:2021wcc}. 
The change in the existing bound on $\kappa_{\tt t}$ in presence of these EFT operators, 
in particular the four fermi operator in context of the present LHC data and at future sensitivities are highlighted in the present work.

Our analysis involves state-of-the-art simulation techniques using both cut-based as well as machine learning (ML) analyses. We point out the key kinematic variables 
that are likely to segregate the signal including EFT operators from that of the SM background contamination. We set limits on the EFT operator coefficients, subject to the 
NP scale, which is likely to be excluded in future luminosities. 

The paper is organized as follows: we discuss the EFT operators that contributes to $thj$ production at LHC in section \ref{sec:EFT}, 
bound on the EFT parameters from existing LHC data in section \ref{sec:bounds}, signal significance and limits at future luminosities in section \ref{sec:reach} and 
conclude in section \ref{sec:summary}.

\section{EFT operators in context of $thj$ production at the LHC}
\label{sec:EFT}

Assuming (as we will) that the NP is not directly observed, all NP effects are parameterized by a series of effective operators. It is well known \cite{Einhorn:2013kja} that there is no 
unique basis of such operators, but that all such bases are related through the equivalence theorem \cite{Arzt:1993gz}; here we choose the so-called Warsaw 
basis \cite{Grzadkowski:2010es}. Though we do not specify the details of the physics underlying the SM, we will assume that it is weakly coupled and decoupling. 

Under these circumstances, it is relevant to differentiate operators according to whether they are generated at 1 or higher loops by the new physics (loop-generated or LG operators), 
or whether they can be generated at tree level (potentially tree-generated or PTG operators) \cite{Einhorn:2013kja}. This separation is useful because the Wilson coefficients of LG 
operators are necessarily suppressed by a factor\footnote{This is the case, for example, for the operator (the notation is explained below \cref{eq:contributing.ops}) $ W_{\mu\nu}^I(\bar q_t \sigma^{\mu\nu} \tau_I u_t) $ whose Wilson coefficient has a natural size of order $g/(4\pi\Lambda)^2$, {\em not} $1/\Lambda^2$.} $ \sim 1/(4\pi)^2 \sim 0.006 $, so that their effects will be generally negligible given the existing experimental precision~\footnote{Ignoring this suppression often leads to the misleading conclusion that the the effects of LG operators can be very important, but such results derive from assuming coefficients that deviate from their natural values by 3 or 4 orders of magnitude. There are, however, a few notable exceptions: SM processes that occur only at one loop, such as $ h \to \gamma\gamma$; and reactions strongly suppressed by small CKM and PMNS matrix elements.}; 
in contrast, PTG operators can have $\mathcal{O}(1) $ Wilson coefficients. It is important to note that LG operators are loop generated by {\em any} kind of NP~\footnote{Assuming 
this NP is described by a local gauge theory containing scalar, fermions and vector fields.}, while PTG operators are generated at the tree level by at least one type of NP, though 
it may be that the NP realized in nature does not share this property. Nonetheless, PTG operators offer the best opportunities for testing in the presence of a wide class of NP.

The SM quark-level amplitude leading to the $thj$ final state are given in Fig. \ref{fig:SM.thj}. 
When NP effects are included, the cross-section contains three terms: the pure SM contribution, 
an NP-SM interference term, and a pure NP term. Contributions involving NP will be suppressed by powers of the NP scale, which we denote by $ \Lambda $; for the case at hand the 
interference term is $ \propto 1/\Lambda^2$ while the pure NP term is $ \propto 1/\Lambda^4 $. In our analysis, we have included both the interference, as well as the pure NP 
contribution. In section \ref{sec:simulation}, we show the contribution of the EFT operators to $thj$ production at different representative benchmark points within the parameter 
space scanned in the analysis. Dimension 8 operators, which also contribute $ \propto 1/\Lambda^4$ via interference with SM,  should, strictly speaking, be included -- we expect, however, that such effects would be of similar order as the ones here obtained and would not change our conclusions; we will neglect them in the following analysis.

We are then interested in those effective operators that are PTG and can interfere with the SM amplitude in the $thj$ process. It follows from the structure of the SM contribution that of the effective operators containing quarks, only those with left-handed light quarks (including the $b$) need be considered. With this in mind the operators we include are:
\bal
\ocal_{\phi\Box} &= |\phi|^2   \Box |\phi|^2\,,\mcr
\ocal_{t\phi} &= |\phi|^2(\bar q_t t_R \phit)  + \text{H.c.}\,,\mcr
\ocal_{qq}\up1 &= (\bar q_t \gamma_\mu q_u)(\bar q_u \gamma^\mu q_t)\,, \mcr
\ocal_{qq;\,1}\up3 &= (\bar q_t \gamma_\mu \tau^I q_t)(\bar q_u \gamma^\mu \tau^I q_u)\,, \mcr
\ocal_{qq;\,2}\up3 &= (\bar q_t \gamma_\mu \tau^I q_u)(\bar q_u \gamma^\mu \tau^I q_t)\,, \mcr
\ocal_{\phi q}\up3 &= (\bar q_t \tau^I \gamma^\mu q_t) (\phi^\dagger i \tau^I D_\mu \phi ) + \text{H.c.}\,.
\label{eq:contributing.ops}
\end{align}
where $ \phi$ denotes the SM scalar doublet; $q_t\,,q_u$ the top-bottom and up-down left-handed quark doublets, respectively; $t_R$ the right-handed top singlet; and $ \tau^I$ the Pauli matrices. We also note that  are two additional 4-fermions operators that contribute to the process of interest and interfere with the SM, but these are Fierz-equivalent to the ones listed (see section \ref{sec:EFT.validity}).

The relevant effective Lagrangian is then
\bal
\lcal_{\tt eff} &= 
\frac{\wc_{\tt h}}{\Lambda^2} \ocal_{\phi\Box} +
\frac{\wc_{\tt t}}{\Lambda^2} \ocal_{u\phi}  +
\inv{\Lambda^2}  \left[ \wc_{qq}\up1 \ocal_{qq}\up1 + \wc_{qq;\,1}\up3 \ocal_{qq;\,1}\up3 + \wc_{qq;\,2}\up3 \ocal_{qq;\,2}\up3 \right]
+ \frac{\wc_{\tt w}}{\Lambda^2}  \ocal_{\phi q}\up3 
+ \text{H.c.} \,;
\mcr
&=\frac{2\wc_{\tt h} \smvev^2}{\Lambda^2} h  \Box h+
\frac{ \wc_{\tt t} \smvev^2}{\sqrt{2}\, \Lambda^2} h (\bar t t )  + \frac{\wc_{\tt 4f}}{\Lambda^2} \left[ (\bar d_L \gamma_\mu u_L) (\bar t_L \gamma^\mu b_L) + \text{H.c.}  \right] \cr
& \qquad\qquad\qquad\qquad +
\frac{\sqrt{2}\, g \wc_{\tt w}  }{\Lambda^2} (h + \smvev)^2 \left( \bar t_L \slashed{W}^+ b_L + \text{H.c.} \right) \,;
\label{eq:leff} 
\end{align}
where we assumed the operator coefficients are real and  
\beq
\wc_{\tt 4f} = \inv3 \wc_{qq}\up1  + 2 \wc_{qq;\,1}\up3   -\inv3 \wc_{qq;\,2}\up3 \,.
\label{eq:def.c4f}
\eeq

A few comments are in order:
\bit
\item The  term $ \propto \wc_{\tt t}$ modifies the top Yukawa coupling: $ m_{\tt t}/\smvev \to (m_{\tt t}/\smvev) \kappa_{\tt t} $ with  $ \kappa_{\tt t} =  1  - \wc_{\tt t} \smvev^3/(\sqrt{2}\,\Lambda^2 m_{\tt t}) $; in the current context the top-quark signal strength for the Higgs is then $ \kappa_{\tt t}   \simeq1- \wc_{\tt t} (250 \, \gev/\Lambda)^2 $. Investigations (see {\it e.g.} \cite{CMS:2018jeh}) that place limits on the value of $ \kappa_{\tt t}$  without considering any other possible NP effects are  {\it de facto} ignoring all but $ \ocal_{t\phi}$ contributions. It is also worth noting that for natural values, $ \wc_{\tt t} = \mathcal{O}(1) $, and $ \Lambda > 1.5\,\tev $, $ |\kappa_{\tt t}-1| \lesssim 0.03 $, which gives the precision required to probe physics beyond the electroweak scale associated with  $ \ocal_{t\phi}$.

\item The  term $ \propto \wc_{\tt w}$  introduces $ Wtb h$ and $ Wtb h^2 $ interactions, and also modifies  the  $Wtb$ coupling, which now becomes $(g/\sqrt{2})(1 + 2 \wc_{\tt w} \smvev^2/\Lambda^2)$; current (3-$\sigma$) limits on the $t$ lifetime then imply $\Lambda/\sqrt{\wc_{\tt w}} > 0.9 $ TeV.

\item We do not include operators that modify the $Wud$ vertex since the current measurements of the hadronic width of the $W$ are precise enough to ensure that such effects, if present, would be unobservable in the $thj$ final state we consider, given the expected precision to which this reaction will be measured. We also ignored all flavor-changing operators and neglected all Yukawa-type interactions but those of the top quark.

\item The  term $ \propto \wc_{\tt h}$ requires a (finite) wave-function renormalization of the Higgs field, $ h \to (1  + \wc_{\tt h} \smvev^2/\Lambda^2) h $, under which all Higgs couplings are modified by the same factor \cite{Einhorn:2013tja}. This operator is generated at tree level by the exchange of either a heavy scalar neutral isosinglet or a scalar isotriplet. For simplicity, in the following we will ignore this operator; our main goal is to show that the introduction of operators other than  $\ocal_{t\phi} $ (as required in an unbiased application of the EFT formalism) can significantly modify the limits on $\kappa_{\tt t}$, and for this it is sufficient to include $\wc_{\tt 4f}$ and $\wc_{\tt w} $ (though, as we will show, the effects of the latter are small). Thus in the discussion below we assume $ \wc_{\tt h} =0$ for simplicity.

\eit

\begin{figure}
$$
\includegraphics[height=5cm]{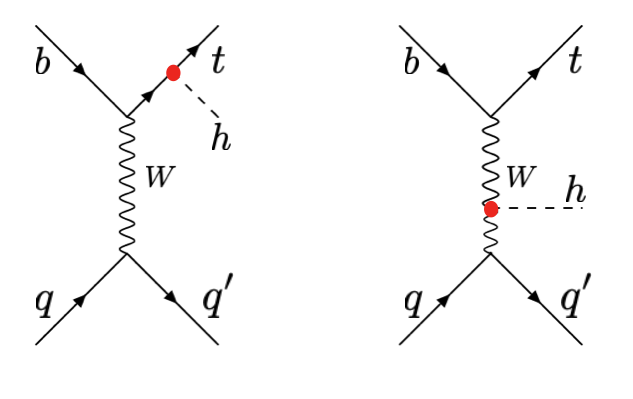}
$$
$$
\includegraphics[height=5cm]{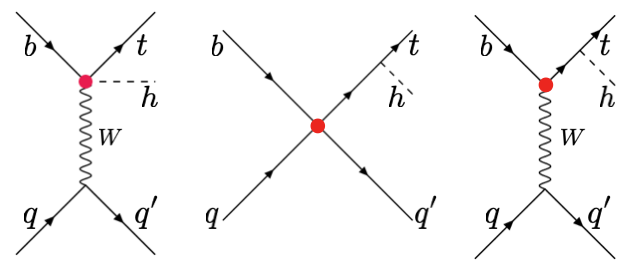}
$$
\vspace{0cm}
\caption{NP-induced vertices (shown by red blob) contributing to single-top production in association with a Higgs at the LHC.}
\label{fig:NP.thj}
\end{figure}

The NP effects we consider modify the $tth$ and $Wtb$ vertices in Fig. \ref{fig:SM.thj} and generate the additional graphs in 
Fig. \ref{fig:NP.thj}. We assume $ \mathcal{O}(1) $ values for the Wilson coefficients and absorb $ |\wc_{\tt w}| $ in the definition of 
$ \Lambda$ (that is, we define the scale of NP of that the coefficient of $ \ocal_{\phi q}\up3 $ is $ \pm 1/\Lambda^2 $ so that our model parameters are
\beq
\text{ NP parameters:}~\{\wc_{\tt 4f} ,\,\wc_{\tt w} = \pm1,  \kappa_{\tt t}, \Lambda\}\,,
\eeq
where we use $ \kappa_{\tt t} $ instead of $ \wc_{\tt t} $ to allow easier comparison with the existing literature.

It is one of the goals of the present investigation to determine the extent to which the presence of non vanishing values of $ \wc_{\tt 4f,\,w}$  affects the determination of $ \kappa_{\tt t} $.

\subsection{Constraints on EFT Operator coefficients}

There are several existing limits on the Wilson coefficients we consider. It is worth noting, however, that these may be conservative as they are often derived by assuming that a 
restricted number of effective operators are present. 

\bit
\item CMS~\cite{CMS:2016uzc} obtains the following  exclusion limit using data from the $7$ and $8$ TeV LHC:
\beq
\frac{\smvev^2}{\Lambda^2} |\wc_{\tt w}| < 0.03 \then  \Lambda  > 1.4\,\tev~~ (\wc_{\tt w} = \mathcal{O}(1))\,.
\eeq
that is comparable to the errors in $|V_{\tt tb}|$ from CMS for the $13$ TeV LHC \cite{CMS:2020vac}. This limit is obtained assuming the presence only 
of an EFT correction to the $Wtb$ vertex, and will be degraded it the  effects of other operators are included.

\item The low energy constraints from $B$ physics \cite{Kozachuk:2020mwa} give
\beq
-0.10 < \frac{\smvev^2}{\Lambda^2} \wc_{\tt w} < 0.15 \then \Lambda > 780 - 640 \,\gev ~~ (\wc_{\tt w} = \mathcal{O}(1))\,.
\eeq

\item The $ t\to bud$ width receives a contribution from $\ocal\up3_{qq}$ and $ \ocal\up3_{\phi q}$:
\beq
\Gamma(t\to bud) = \frac{m_{\tt t} g^4 I_1 }{2(8\pi)^3} \left[  V_{\tt tb}^2 + 2 b_{\tt r} + 2 a_{\tt r} \frac{I_2}{I_1} - 2 \gamma  a_{\tt i}  \right] + \mathcal{O}(1/\Lambda^4)\,,
\eeq
where $g$ is the $\su2_L$ gauge coupling, $ \gamma = \Gamma_{\tt w}/m_{\tt w}\simeq0.026,\,r = m_{\tt w}/m_{\tt t} \simeq 0.46$, 
\beq
a_{\tt r} + i a_{\tt i} = \frac{2 \smvev^2 }{\Lambda^2} \wc_{\tt 4f}\,; \qquad 
b_{\tt r} = \frac{2 \smvev^2}{\Lambda^2} \re \wc_{\tt w}\,;
\eeq
and~\footnote{To $O(1/\Lambda^2)$ the width does not depend on $\im \wc_{\tt w}$.}
\bal
I_1 &= \inv{\gamma r^2} \left\{ \pi + \arctan \left[\frac\gamma{(1+\gamma^2)r^2 - 1} \right]\right\} \simeq 553.6\,, \mcr
I_2 &= \inv{2r^2} \ln \left[ \frac{(1-1/r^2)^2 + \gamma^2}{1+ \gamma^2} \right] \simeq 5.9\,.
\end{align}
The sensitivity to $a_{\tt i} $ is much diminished because it appears multiplied by $ \gamma$, the sensitivity to $ a_{\tt r}$ is also small because $I_1 \gg I_2 $ ($I_1$ gives the leading $ \propto 1/\gamma$ contribution in the limit $ \gamma \to 0 $). The $3\sigma$ error in the width is about 40\%; applying this to each coefficient separately~\footnote{These limits do not hold if there are cancellations.}
\beq
|a_{\tt i}| < 7\,,\quad |a_{\tt r}|<19\,, \quad |b_{\tt r}|<0.2\,.
\eeq
For $\mathcal{O}(1)$ coefficients this gives $ \Lambda > 130\,\gev,\,80\,\gev,\,780\,\gev$ respectively.

\item Higgs production cross-section via gluon fusion is sensitive to $\wc_{\tt t}$ and $ \wc_{\tt h}$. The contribution from these effective operators gives \cite{Einhorn:2013tja}
\beq
\sigma_{gg\to\tt h} = \left( 1 +  \frac{\sqrt{2}\smvev^3}{m_{\tt t}\, \Lambda^2} \wc_{\tt t} +  \frac{2\smvev^2}{\Lambda^2} \wc_{\tt h} \right) \sigma_{gg\to\tt h} \up{\tt SM}\,,
\eeq
where we assumed $ \wc_{\tt t} $ real. The $ 3 \sigma $ error on this cross-section is $ \sim 20\% $ that translates into $ \Lambda > 780\,\gev $ for $ \wc_{\tt t,\,h} = \mathcal{O}(1)$ (and no cancellations).

\item The coefficient $ \wc_{\tt h} $ is also constrained by the $ h \to ZZ^*,\,WW^*$ decay widths \cite{Einhorn:2013tja}:
\beq
\frac{\Gamma(h\to ZZ^*)}{\Gamma_{\tt SM}(h\to ZZ^*)} = \frac{\Gamma(h\to WW^*)}{\Gamma_{\tt SM}(h\to WW^*)} = 1 + \frac{v^2}{\Lambda^2} \wc_{\tt h}\,.
\eeq
The $ZZ^*$ and $WW^*$ widths have $3\sigma$ uncertainties of $\sim25\%$ and $\sim30\%$, respectively, which give $\Lambda > 500 \,\gev,\, 450\, \gev $, when $ |\wc_{\tt h}| \sim 1 $.

\item The TopFitter collaboration \cite{Buckley:2015lku} reports the individual limits, 
\bal
\Lambda > 960\,\gev \quad &\text{when} ~ |C_{\tt4f}|=1\mcr
\Lambda > 720\,\gev \quad &\text{when} ~ |C_{\tt w}|=1
\end{align}
at 95\% CL (the marginalized limits are $800\,\gev$ and $510\,\gev$, respectively); these are comparable to the above constraints.

\eit

\subsection{Validity of the EFT approximation}
\label{sec:EFT.validity}

Under the current assumptions of a weakly-coupled and decoupling heavy physics, the fundamental limitation in using an EFT approximation is the 
condition that the typical energy is smaller than $ \Lambda $, for otherwise, the experiment(s) under consideration would have sufficient energy to 
directly create the new heavy particles. However, in order to impose this condition, the definition of `typical' energy requires some clarification.

Consider, for example, the 4-fermion vertex $(\bar d_L \gamma_\mu u_L)(\bar t_L \gamma^\mu b_L)$ derived from the 4-fermion operators in 
\cref{eq:contributing.ops}. As noted in that section, there are other operators that generate the same vertex, but these are Fierz-equivalent to 
those in \cref{eq:contributing.ops}; explicitly
\bal
\ocal_{qq}\up{8;\,1} &= (\bar q_t \gamma_\mu T^A  q_u)(\bar q_u \gamma^\mu T^A  q_t) = \frac23 \ocal_{qq}\up1 - \ocal_{qq;\,1}\up3 - \ocal'  \,,\mcr
\ocal_{qq}\up{8;\,3} &= (\bar q_t \gamma_\mu T^A \tau^I q_u)(\bar q_u \gamma^\mu T^A \tau^I q_t) = \frac23  \ocal_{qq;\,2}\up3 +  \ocal_{qq;\,1}\up3 - \ocal' \,,
\label{eq:other.ops}
\end{align}
where $ \ocal' = (\bar q_t \gamma_\mu  q_t)(\bar q_u \gamma^\mu  q_u) $. 

It is worth noting that neither $ \ocal' $ nor the two related operators $ (\bar q_t \gamma_\mu T^A \tau^I q_t)(\bar q_u \gamma^\mu T^A \tau^I q_u) $ and $ (\bar q_t \gamma_\mu T^A \tau^I q_t)(\bar q_u \gamma^\mu T^A \tau^I q_u) $ produce a vertex $(\bar d_L \gamma_\mu u_L)(\bar t_L \gamma^\mu b_L)$, and do not interfere with the SM contribution; they are not discussed further for this reason.

From these considerations we can immediately determine the type of heavy physics that can generate the vertex $(\bar d_L \gamma_\mu u_L)(\bar t_L \gamma^\mu b_L)$ at tree level:

\beq
\begin{array}{c|c|c|l|l|l}
\text{vector}	& \su2 		& \su3 & \text{couplings} & \text{operator generated} & \text{reference}\cr\hline
X\up{1;1}				& {\bf1}	& {\bf1}	& \bar q_u \gamma^\mu q_t					&\ocal_{qq}\up1       &\text{\quad\cref{eq:contributing.ops}} \cr
Y_1\up{3;1}				& {\bf3}	& {\bf1}	& \bar q_a \gamma^\mu \tau^I q_a,~(a=u,t)~~&\ocal_{qq;\,1}\up3   &\text{\quad\cref{eq:contributing.ops}} \cr
Y_2\up{3;1}				& {\bf3}	& {\bf1}	& \bar q_u \gamma^\mu \tau^I q_t			&\ocal_{qq;\,2}\up3   &\text{\quad\cref{eq:contributing.ops}} \cr
Y_3\up{3;8}			& {\bf3}	& {\bf8}	& \bar q_u \gamma^\mu \tau^I T^A q_t		&\ocal_{qq}\up{8;\,3} &\text{\quad\cref{eq:other.ops}}		 \cr
V\up{1;8}				& {\bf1}	& {\bf8}	& \bar q_u \gamma^\mu T^A q_t				&\ocal_{qq}\up{8;\,1} &\text{\quad\cref{eq:other.ops}}		 \cr
\end{array}
\label{mediators}
\eeq
(superscripts in the vector fields correspond to their $\su2\times\su3$ transformation properties). Consider now the reactions\footnote{$t \bar b$ can contribute to the final state of interest when $b$ is mistagged as a forward light jet.} $ b q\to t q'  $ and $ q q'\to t \bar b $, where $q,\,q'$ are light quarks or antiquarks, then the above heavy vectors contribute in either s or t channels as listed below:
\beq
\begin{array}{c|ccccc}
				& X\up{1;1}	& V\up{1;8}	& Y_1\up{3;1}	& Y_2\up{3;1}	& Y_3\up{3;8}	\cr \hline
b q \to t q'	& \text{s}	& \text{s}	& \text{t}		& \text{s}		& \text{s}		\cr
qq'\to t \bar b	& \text{t}	& \text{t}	& \text{s}		& \text{t}		& \text{t}		\cr
\end{array}
\label{s-t}
\eeq

For the process $ q q'\to t \bar b $ with a quark-level CM energy $ E_{\tt CM}(t\bar b) $ the EFT will be valid provided the mass of the vector boson obeys $ m(Y_1) >  E_{\tt CM}(t\bar b) $;  
the corresponding constraints for the other vector boson masses will be much weaker because the typical values of the t-channel momentum transfer are much smaller than 
$ E_{\tt CM}(t\bar b) $.  In contrast, for the process $ b q\to t q'  $ we must have $m(X),\,m(V),\, m(Y_2)$ and $ m(Y_3) $ all larger than $ E_{\tt CM}(t q') $ (the CM energy for 
that reaction) for the EFT approximation to be valid. Therefore the potential NP contribution to $tq'$ production comes from $Y_1\up{3;1}$, which has subdued contribution 
to $t\bar{b}$ production, and is neglected. Experimental limits (see \cite{Stamm:2018wrx} and references therein) requires $ m(Y_1) $ to be above several 
TeV~\footnote{We may also note here that in SM $q q'\to t \bar b$ production is small, for example, $t h\bar{b}$ production cross-section is $2.26$ fb at the 
LHC with $\sqrt s=14\,\tev$, significantly below than $thj$ production cross-section.}. The corresponding analysis for the reaction at hand is discussed in \cref{sec:simulation} below.

\section{Limits using current LHC data}
\label{sec:bounds}

The essence of studying EFT operator contributions to the SM processes at collider is dominantly two fold: $(i)$ estimating the limit on NP parameters 
for current CM energy and luminosity, and $(ii)$ evaluating the contribution of EFT operators to future sensitivities to find out the discovery limit 
after a careful background estimation/reduction. The methodology for both are similar; however in the first case, one needs to adhere to the 
event selection strategy used in the existing experimental analysis as closely as possible. 
We elaborate upon the first part in this section, while the discovery
 potential in future luminosities is discussed in the next section. 

\subsection{Model implementation and production cross-section}

\begin{figure}[htb!]
$$
\includegraphics[height=5.0cm]{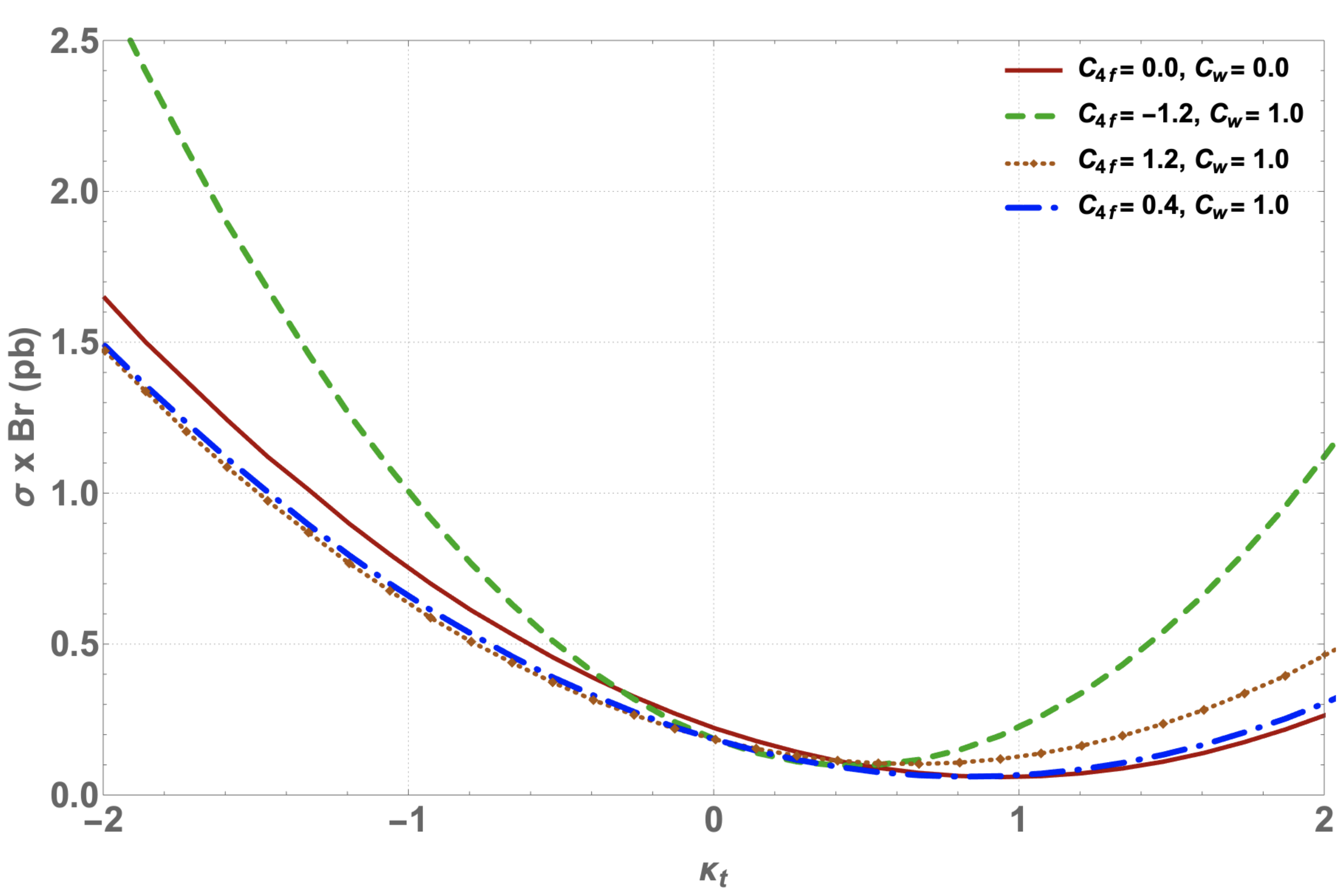}
\includegraphics[height=5.0cm]{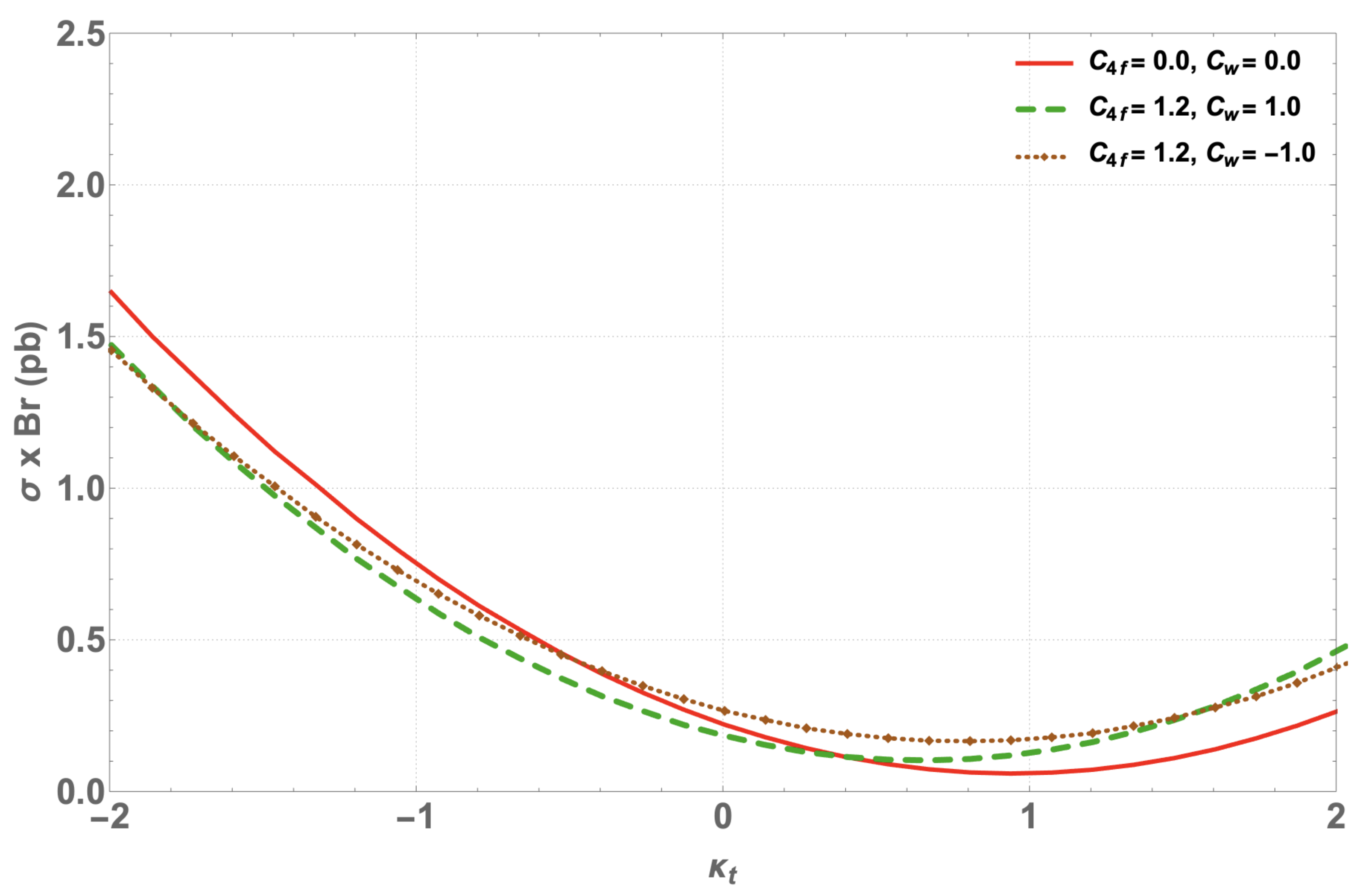}
$$
\caption{Dependence of $thj$ production cross-section on $ \kappa_{\tt t} $ in presence of EFT operators when $ \Lambda =1 \, \tev $. Left: $\wc_{\tt w} = 1 $ 
and several values of $ \wc_{\tt 4f}$ chosen as $\{-1.2,+1.2, 0.4\}$; Right: $\wc_{\tt4f}=1$ and $\wc_{\tt w}$ are chosen as $\{-1,1\}$. We also compare them 
with the case where $ \wc_{\tt 4f}=  \wc_{\tt w}=0$, with variable $\kappa_{\tt t}$ (shown by the red thick line). All cross-sections are evaluated at the LHC with $\sqrt s=13$ TeV. 
The branching ratio (Br) of Higgs includes $h \to WW^*,ZZ^*, \gamma\gamma, b\bar{b}, \tau\bar{\tau}$.}
\label{fig:EFT-cs}
\end {figure}

We first estimate the contribution of EFT operators to $thj$ production following the operators listed in Eq.~\eqref{eq:leff} 
and Feynman graphs as in \cref{fig:NP.thj}. The methodology is simple. We insert the effective Lagrangian Eq.~\eqref{eq:leff} in the 
 {\tt FeynRules} \cite{Alloul:2013bka} code and create files compatible for event simulation in {\tt MadGraph5\_aMC@NLO} \cite{Alwall:2011uj}. 
 We generate events including SM and EFT contributions together, and plot the cross-section as a function of $\kappa_{\tt t}$ 
 (which is equivalent to $ \wc_{\tt t}$) at $\sqrt s=$13 TeV (we use this notation to simplify the comparison of the outcome with CMS results in \cite{CMS:2018jeh}). 
In \cref{fig:EFT-cs}, we show the variation of $thj$ cross-section 
 times the branching ratio (Br) of Higgs $ \to WW^*,ZZ^*, \gamma\gamma, b\bar{b}, \tau\bar{\tau}$, in presence of EFT operators 
assuming the NP scale $\Lambda=1$ TeV. In the left plot we have kept fixed $\wc_{\tt w}=1$ and chosen different $\wc_{\tt4f}$ as $\{-1.2,\,+1.2, 0.4\}$ ({\it cf.} the figure inset). 
On the right panel, a similar calculation is done fixing $\wc_{\tt4f}=1$ and varying $\wc_{\tt w}$ for two values $\{-1,\,1\}$. 
All these points are compared to the case where both $\wc_{\tt w}, \wc_{\tt 4f}=0$, keeping $\kappa_{\tt t}$ as variable on which the cross-section depends.
In both the cases, we see that with larger $|\kappa_{\tt t}|$, the cross-section grows, as expected. Importantly, it is worth noting that for $\wc_{\tt4f}>0$, the $thj$ cross-section is 
smaller than SM only contribution for $\kappa_{\tt t}<0$. Similarly, for $\wc_{\tt4f}<0$, the cross-section is larger than the SM only contribution 
for both positive and negative $\kappa_{\tt t}$. We also can see that the contribution of $\wc_{\tt w}$ is much milder than that of $\wc_{\tt4f}$. 
It is worth reminding here that in the recent analysis for single top production at LHC \cite{Degrande:2018fog}, the contribution from $\wc_{\tt4f}$ has 
completely been ignored, which clearly plays a more dominating role than other operators.

\begin{figure}[htb!]
$$
\includegraphics[height=7.0cm]{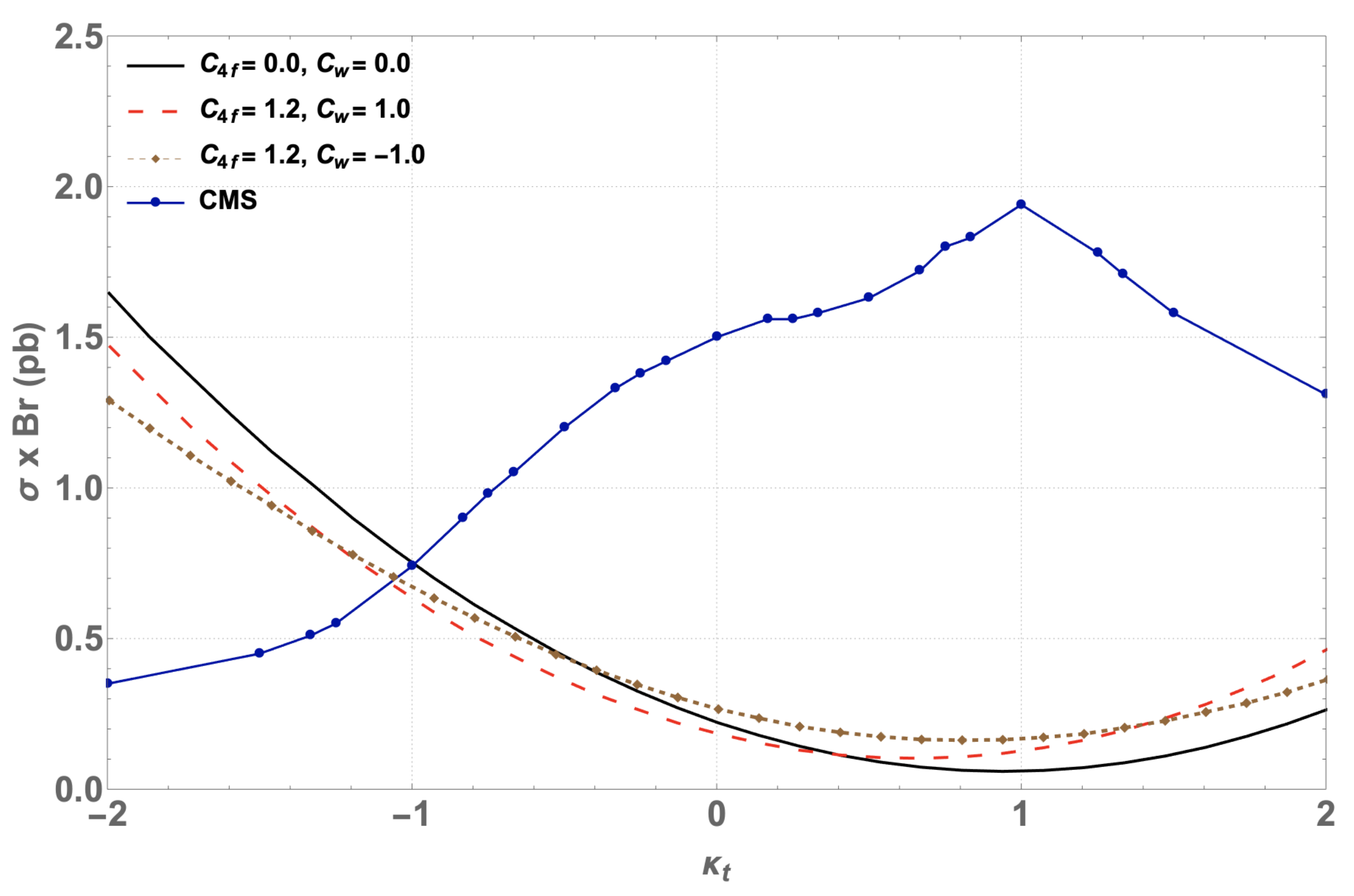}
$$
\caption{$thj$ cross-section ($\sigma$) times branching ratio (Br) of Higgs $ \to WW^*,ZZ^*, \gamma\gamma, b\bar{b}, \tau\bar{\tau}.$
for several benchmark points of the model (see figure legend), including 
that of $ \wc_{\tt 4f},  \wc_{\tt w}=0$ and variable $\kappa_{\tt t}$ (black thick line), confronted with CMS observed data (dotted solid line) \cite{CMS:2018jeh}. Note that the data (dotted solid line) meets the SM cross-section (black solid line) 
at $\kappa_{\tt t}= -0.9$.}
\label{fig:EFT-bound}
\end {figure}

The comparison to the recent most LHC data \cite{CMS:2018jeh} is shown in \cref{fig:EFT-bound}. In this figure, along the y-axis here, we plot 
$thj$ production cross-section times the branching ratio of Higgs to all final states except for the gluon pair 
($h \to WW^*,ZZ^*, \gamma\gamma, b\bar{b}, \tau\bar{\tau}$), in accordance to the CMS analysis \cite{CMS:2018jeh}. 
The dot-thick line in blue indicates the bound obtained from the LHC data with 
$\sqrt{s}=13$ TeV and integrated luminosity $\mathcal{L}=35.9$ fb$^{-1}$ as a function of $\kappa_{\tt t}$. 
For model example, we chose combinations of Wilson coefficients that provide 
good illustrations of the important operators other than $ \ocal_{t\phi}$ can have (see figure inset legends); 
for example choosing $\wc_{\tt w}=1,~\wc_{\tt 4f}=1.2$ minimizes the cross-section 
when $ \kappa_{\tt t}=-1$, while $\wc_{\tt w}=-1.0,~\wc_{\tt 4f}=0.8$ corresponds to maximum cross-section when $\kappa_{\tt t}=+1$, when the Wilson coefficients are 
varied within $\{-1.2,\,+1.2\}$. The plot shows that the presence of $\ocal\up3_{qq,\,\phi q}$ can alter the limits  $\kappa_{\tt t}< -0.9$ 
obtained  \cite{CMS:2018jeh} in absence of these operators. We can also see that for $\wc_{\tt w}=\wc_{\tt 4f}=0$ and $ \kappa_{\tt t}=1$, we recover the SM cross-section 
times branching ratio $\sim$ 70 fb.

There are a few more comments in order. First, although we clearly see that $ \wc_{\tt h,\,4f,\,w} \not=0$ changes the experimental limit as
in \cref{fig:EFT-bound}, one needs to set a criteria for obtaining the bound on the EFT parameters from the existing data. 
This is elaborated on in the next subsection. Second, the presence of these EFT operators provides a significant departure from the signal 
events compared to the SM, so judicious cuts on kinematic variables or BDT analysis paves the way for discovery potential. 
Lastly, we note that these operators contribute to other processes like top decay, etc.; while choosing 
the values of the Wilson coefficients and NP scale, we abide by such limits.  

 Before proceeding further we note here that EFT approximation at collider is valid when the NP scale is larger than the CM energy of the reaction,
\bea
\rm {Validity ~ of~EFT~at~LHC}:~ \Lambda > \sqrt{\hat{s}}\,.
\eea

\begin{figure}[htb!]
$$
\includegraphics[height=6.0cm]{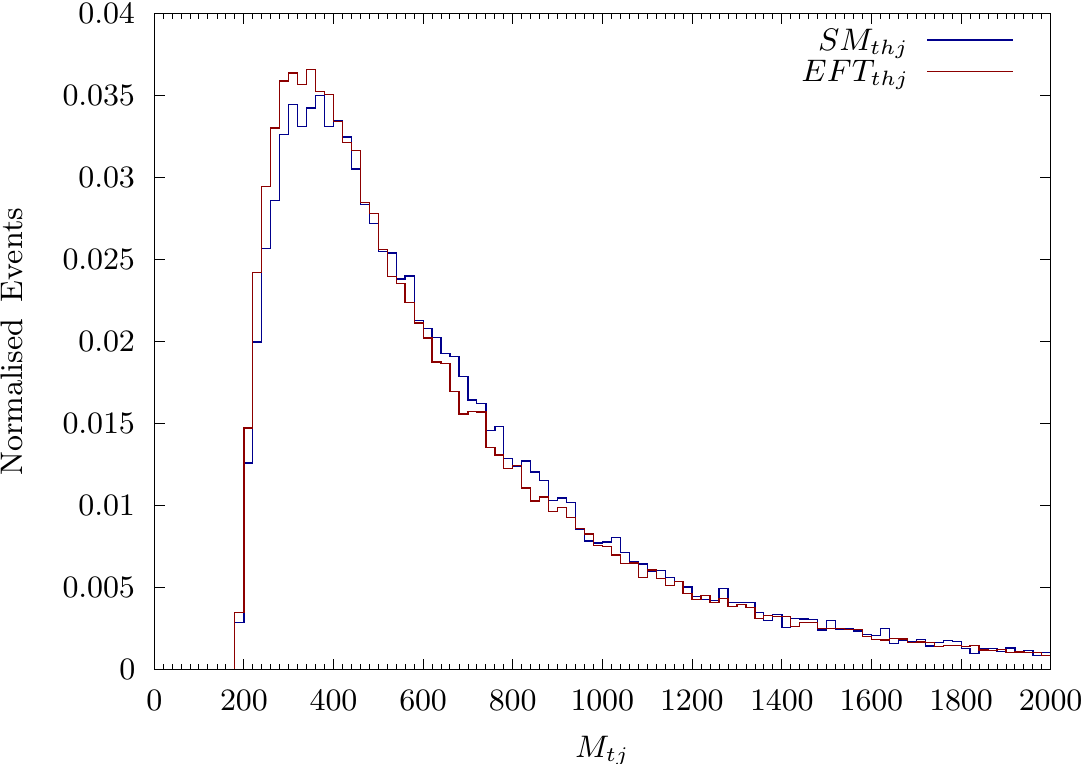}
$$
\caption{Invariant mass ($M_{tj}$) distribution for the $thj$ signal production with EFT contributions assuming 
$\Lambda=1$ TeV and all the parameters $\{\wc_{\tt 4f} ,\,\wc_{\tt w},  \kappa_{\tt t}\}$ set to one, at LHC with $\sqrt{s}=13$ TeV. 
We also depict the distribution with SM contribution only. The production cross-section is normalized to one.}
\label{fig:minv}
\end {figure}

However, ensuring this limit is difficult at LHC, due to the unknown and variable subprocess CM energy $\hat{s}$. 
One can however have an estimate of the partonic CM energy by constructing the invariant mass of the final state particles,
\bea
M_{inv}^2=\left(\sum_i p_i \right)^2\,;
\eea
where $p_i$ represents four-momenta and $i$ runs over all or a subset of the visible set of detectable final state particles. 
The appropriate choice of invariant mass depends on whether it is a $s/u/t$ channel reaction, 
thus segregating different types of NP contributions, see for example, \cite{Bhattacharya:2015vja}. Now, if the $M_{inv}$ distribution 
peaks at a smaller value than the chosen NP scale ($\Lambda$), then the bulk of the events are produced with CM energy smaller than $\Lambda$, 
validating the EFT nature of the interaction. In \cref{fig:minv}, we plot invariant mass distribution $M_{t j}$ 
for SM plus EFT contribution assuming $\Lambda=1$ TeV, with $\{\wc_{\tt 4f} ,\,\wc_{\tt w},  \kappa_{\tt t}\}=1$ at the LHC with $\sqrt{s}=13$ TeV, 
to show that the peak arises $\sim 500$ GeV $<\Lambda$, thus validating of the EFT nature of the interaction.

\subsection{Event Analysis and simulation techniques}
\label{sec:simulation}

We generate all the event samples using { \tt MadGraph5\_aMC@NLO } \cite{Alwall:2011uj} at parton-level in leading order (LO) and a dedicated universal 
FeynRules output (UFO) model for the EFT framework was produced using {\tt FeynRules}. We have used the 5-flavor scheme \cite{Maltoni:2012pa} of 
parton distribution function with the NNPDF30LO PDF set \cite{NNPDF:2014otw}. The default  {\tt MadGraph5\_aMC@NLO} LO dynamical scale was used, 
which is the transverse mass calculated by a $k_{T}$-clustering of the final-state partons. Then we interface the events with the Pythia 8 \cite{Mrenna:2016sih} 
parton shower. Events of different jet-multiplicities were matched using the MLM scheme \cite{Mangano:2006rw} with the default {\tt MadGraph5\_aMC@NLO} 
parameters and all samples were processed through {\tt Delphes 3} \cite{deFavereau:2013fsa}, which simulates the detector effects, applies simplified 
reconstruction algorithms and was used for the reconstruction of electrons, muons, and hadronic jets. 

For the leptons $\ell$ (electrons and muons) the reconstruction was based on transverse momentum ($p_{T}$)-and pseudo-rapidity ($\eta$)-dependent efficiency 
parametrization. Isolation of a lepton from other energy-flow objects was applied in a cone of $\Delta R =\sqrt {\Delta \eta^2+\Delta \phi^2}$ = 0.4 with a minimum 
$(p_T)_\ell >$ 25 GeV. We reconstruct the jets using the anti-$K_T$ clustering algorithm \cite{Cacciari:2008gp} with a radius parameter of $\Delta R = 0.4$ 
implemented in {\tt FastJet} \cite{Cacciari:2011ma,Cacciari:2005hq}. The identification of $b$-tagged jets was done 
by applying a $p_{T}$-dependent weight based on the jet’s associated flavor, and the {\tt MV2c20} tagging algorithm \cite{ATLAS:2015dex} in the 70\% working point.

$thj$ production within SM and beyond (including EFT operators) leads to different possible final states following subsequent 
decay modes of the top quark and the Higgs boson. Each different final state faces different non-interfering SM background contributions, 
thereby altering the signal extraction strategy and event simulation. We focus here on the same sign dilepton (SSD) signal with both leptons 
having positive or negative charges that may arise with different flavours ($\ell=e,\mu$) associated with one forward jet \footnote{By forward jet, we refer to the jet having 
maximum pseudorapidity in an event.} and two 
other jets ($j$), resulting from the following decay chain:
\beq
\nonumber {\rm Signal~({\bf SSD})}: pp \to thj, ~t\to bW^+\to b\ell^+\nu, ~h\to W^-W^{+*} \to (jj)(\ell^+\nu).
\eeq 

The dominant SM background arises from $t\bar{t}W$, which also contributes to the same final state signal. Other processes that 
contribute as non-interfering background are $t\bar{t}Z$, $WZ$ and $t\bar{t}$. The contribution from $t\bar{t}$ to the same sign dilepton signal
arises due to the leptons coming out of bottom meson decay. Apart, $t\bar{t}h$ can also contribute to the signal, which is sensitive to the 
Wilson coefficients, $\kappa_{\tt t}$ in particular. As we are not tagging the top quarks in the final state, $t\bar{t}h$ can be 
considered as part of the signal in principle \footnote{The CMS analysis \cite{CMS:2018jeh} also has considered $t\bar{t}h $ as part of the signal.}. 
However, as a conservative estimate, we do not consider this process while scanning the parameter space for the signal significance, 
although the contribution at the selected benchmark points are estimated.
Adhering to the CMS analysis \cite{CMS:2018jeh}, we adopt the basic cuts as follows: 
\begin{itemize}
\item Demanding exactly two same sign dilepton ($\ell$) with the combination $\mu^\pm\mu^\pm$ or $e^\pm\mu^\pm$, following the reconstruction 
methodology as described above. 
\item $(p_T)_\ell>25/15$ GeV and no loose lepton with invariant mass $m_{\ell\ell}<12$ GeV.
\item Event should comprise of one or more $b-$tagged jet with $p_T>$ 25 GeV within $|\eta|<2.4$.
\item One or more untagged jets with $p_T>$ 25 GeV within $|\eta|<2.4$, and $p_T>$ 40 GeV for $|\eta|>2.4$ following the jet reconstruction technique.
\end{itemize}

We further define, 
\bea
\epsilon_{cut}=\frac{\sigma^{\rm SSD}}{\sigma^{\rm prod}} \,,
\eea  
where $\sigma^{\rm SSD}$ indicate final state SSD signal cross-section surviving after taking into account 
appropriate branching ratios and basic cuts and  $\sigma^{\rm prod}$ represents $thj$ production cross-section. 

As the observed number of events at the LHC after the final BDT analysis is not available, it is important to ask how 
to set a limit on our EFT parameter space $\{\wc_{\tt 4f} ,\,\wc_{\tt w},  \kappa_{\tt t}, \Lambda\}$ from the observed data. 
Following the CMS observation  \cite{CMS:2018jeh}, we simulate $thj$ production for SM processes {\it without} EFT contribution for
$\kappa_{\tt t} = -0.9$ to produce same sign dilepton (SSD) events using the basic cuts at $\wc_{\tt 4f}$
$\sqrt s=13$ TeV and integrated luminosity $\mathcal{L}=35.9$ fb$^{-1}$. Thus obtained number of events ($N_0$) is used 
to set the limit on the EFT parameters,

\bea
N_0= (\sigma^{thj})_{SM}\big{\vert}_{\kappa_{\tt t}=-0.9}\times \epsilon_{cut}~(0.15 \%) \times \mathcal{L} ~(35.9 ~\rm{fb}^{-1})= 37 \,.
\label{bound}
\eea

We note here that $N_0$ is specific to the SSD channel produced from $thj$, which uses the CMS exclusion limit of 
$\kappa_{\tt t}$ inferred from all possible Higgs decay modes excepting for gluon gluon. Also there is a certain element of uncertainty 
in the estimation of this number due to the model dependence of the $\epsilon_{cut}$, which is rather small given the range of Wilson coefficients 
scanned in this analysis, as explained later on.

\begin{table}[htb]
\begin{center}
\begin{tabular}{|c|c|c|c|c|}
\hline
\hline
Benchmarks & \{$\kappa_{\tt t}, \wc_{\tt 4f}, \wc_{\tt w}$\} & $\sigma_{\rm prod}$ (fb) & $N_{bc}$ & $\epsilon_{cut}$ \cr 
\hline
\hline
SM $thj$ & \{1.0, 0, 0\} & 71.0 & 3.9 & 0.0015 \cr
\hline
BP0 &  \{-1.0, 0, 0\} & 888.81 & 47.61 & 0.0017\cr
\hline
BP1 & \{-1.0,1.2,1.0\} & 773.05 & 39.0 & 0.00172 \cr
\hline
BP2 & \{-0.5, 1.2,-1.0\}  &  522.29 & 26.46 & 0.001675 \cr
\hline
BP3 & \{0.5, 0.4, 1.0\} & 95.06 & 4.38 & 0.00152 \cr
\hline
BP4 & \{1.0, 0.8, -1.0\}  & 134.05 & 6.01 & 0.0014 \cr
\hline
\hline
$t\bar{t}W$ & \{1.0, 0, 0\} & 566 & 164.14 & 0.013 \cr
\hline
$t\bar{t}Z$ & \{1.0, 0, 0\} & 863 & 78.39 & 0.013  \cr
\hline
$t\bar{t}$ &  \{1.0, 0, 0\} & 839 $\times$ $10^{3}$ & 198.17 & 1.6 $\times$ $10^{-5}$ \cr
\hline
\hline
\end{tabular}
\caption{Production cross-section ($\sigma_p$), number of SSD signal events after the basic cuts ($N_{bc}$) and cut efficiency ($\epsilon_{cut}$) 
in SM, together with EFT contributions for benchmark points BP0-BP4 at the LHC with $\sqrt s=13$ TeV and integrated luminosity $\mathcal{L}=35.9$ fb$^{-1}$. 
SM background ($t\bar{t}W$, $t\bar{t}Z$, $t\bar{t}$) contributions are also noted after including appropriate $K$ factors (see text). 
All the cross-sections for the signal benchmark points are obtained after multiplying by the $K$ factor 1.187. }
\label{tab:basic cuts} 
\end{center}
\end{table}

\begin{table}[htb]
\begin{center}
\begin{tabular}{|c|c|c|c|}
\hline
\hline
\text{Model} &          \text{Contribution with $\wc_{\tt 4f}=\wc_{\tt w}=0$}         &             \text{Interference contribution} &                          \text{Pure EFT contribution}\cr \hline
BP1 	& 748.6	 &     -221.38 $\left(1 ~\rm TeV/\Lambda\right)^2$      &        123.78 $\left(1 ~\rm TeV/\Lambda\right)^4$   \cr \hline
BP2 & 435.0	&     -24.43 $\left(1 ~\rm TeV/\Lambda\right)^2$         &        28.79 $\left(1 ~\rm TeV/\Lambda\right)^4$ \cr \hline
BP3 	& 91.52	&	-21.67 $\left(1 ~\rm TeV/\Lambda\right)^2$ 	&	        9.38 $\left(1 ~\rm TeV/\Lambda\right)^4$ \cr \hline
BP4 	& 59.8 	&	-18.32 $\left(1 ~\rm TeV/\Lambda\right)^2$        &		        71.42 $\left(1 ~\rm TeV/\Lambda\right)^4$ \cr \hline
\hline
\end{tabular}
\caption{Contribution to $thj$ production cross-section (in fb) from the dimension six EFT operators considered in Eq.~\eqref{eq:contributing.ops}, 
divided into interference term ($\sim 1/{\Lambda^2}$) and pure NP contribution ($\sim 1/{\Lambda^4}$) with $\Lambda$ in TeV for the 
benchmark points as in Table~\ref{tab:basic cuts}. One can obtain the cross-sections as in Table~\ref{tab:basic cuts} using $\Lambda=1$ TeV after multiplying by the 
$K$ factor 1.187.}
\label{tab:EFT-contribution}
\end{center}
\end{table}

We next choose a few representative benchmark points with EFT contributions (BP0-BP4) and note the production cross-section ($\sigma_p$),
number of events after the basic cuts ($N_{bc}$) and cut efficiency $\epsilon_{cut}$ at the LHC with $\sqrt s=13$ TeV and $\mathcal{L}=35.9$ fb$^{-1}$ 
in Table \ref{tab:basic cuts}. SM $thj$ production, NLO contributions from the dominant SM backgrounds $t\bar{t}W$, $t\bar{t}Z$, along with $t\bar{t}$ are all pointed out.
The $K$- factors used for $t\bar{t}W (t\bar{t}Z)$ is 1.78 (1.51) \cite{CMS:2017ugv}, SM $thj$ is 1.18 \cite{CMS:2018jeh}.

The benchmark points are chosen with different values of $\kappa_{\tt t}$ with a combination of 
\{$\wc_{\tt 4f}, \wc_{\tt w}$\} so that they capture some interesting physics, where the Wilson coefficients vary maximally within $\{-1.2,1.2\}$. 
For example, BP1: $\{\kappa_{t}, \wc_{\tt 4f}, \wc_{\tt w}\}=\{-1.0, 1.2,1.0\}$ {\it minimizes} the $thj$ cross-section at $\kappa_{t}=-1$
and BP4: $\{1.0, 0.8,-1.0\}$ {\it maximizes} the cross-section at $\kappa_{t}=+1$, within the range in which the Wilson coefficients have been varied. 
BP2: $\{-0.5, 1.2,-1.0\}$ and BP3: $\{0.5, 0.4,1.0\}$ are chosen with intermediate $\kappa_{\tt t}$ 
\footnote{For negative (positive) $\kappa_{\tt t}$, similar choice of \{$\wc_{\tt 4f}, \wc_{\tt w}$\} minimizes/maximizes the cross-section.}. 
BP0 refers to a point with $\kappa_{\tt t}=-1$ with other NP couplings set to zero; when compared to BP1, this shows how the choices of  
$\{\wc_{\tt 4f}, \wc_{\tt w}\}$ can alter the cross-section for a given $\kappa_{\tt t}$. Also note that BP1 produces events very close to the exclusion 
limit and kept as a reference due to uncertainties in pdfs, choice of renormalization, and factorization scales at the hadron collider. 
In Table~\ref{tab:EFT-contribution}, the relative contribution of the interference term and pure EFT contribution to the $thj$ production cross-section
at the benchmark points are shown as a function of $\Lambda$ (TeV). We see that when the pure $\kappa_{\tt t}$ dependent contribution is small 
(this can happen when $\kappa_{\tt t}$ takes positive values or small negative values) and/or $\wc_{\tt 4f}$ is large, pure EFT contribution $\sim 1/{\Lambda^4}$ 
can be significant. We would also like to mention that the benchmark points presented here are just indicative 
of different features of the EFT contribution to $thj$ signal, while a detailed scan in the 
$\kappa_{\tt t},\wc_{\tt 4f}$ plane is provided both in the current limit and for future sensitivities of LHC for different choices of $\wc_{\tt w}$.

\begin{figure}[htb!]
$$
\includegraphics[height=7.0cm]{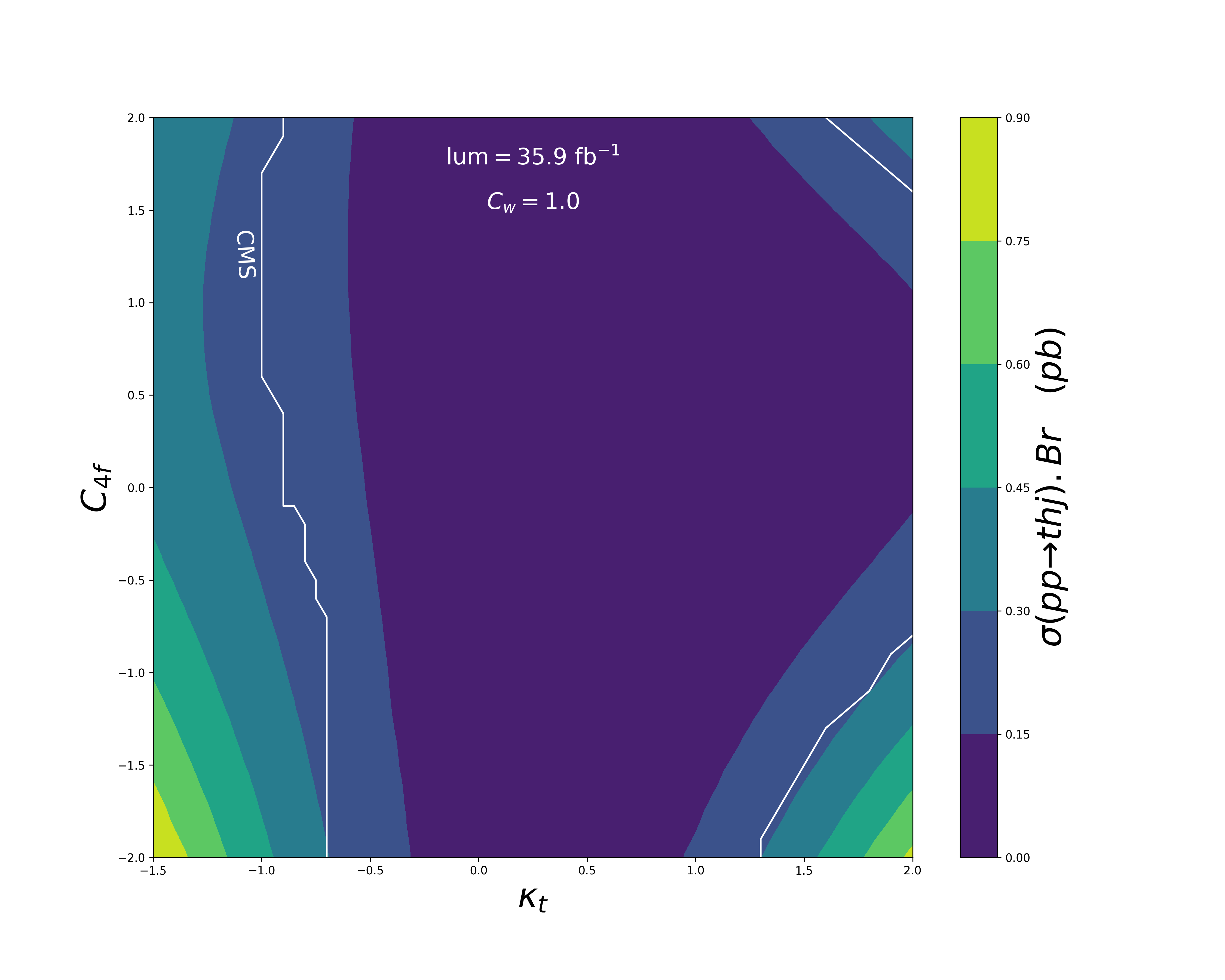}
\includegraphics[height=7.0cm]{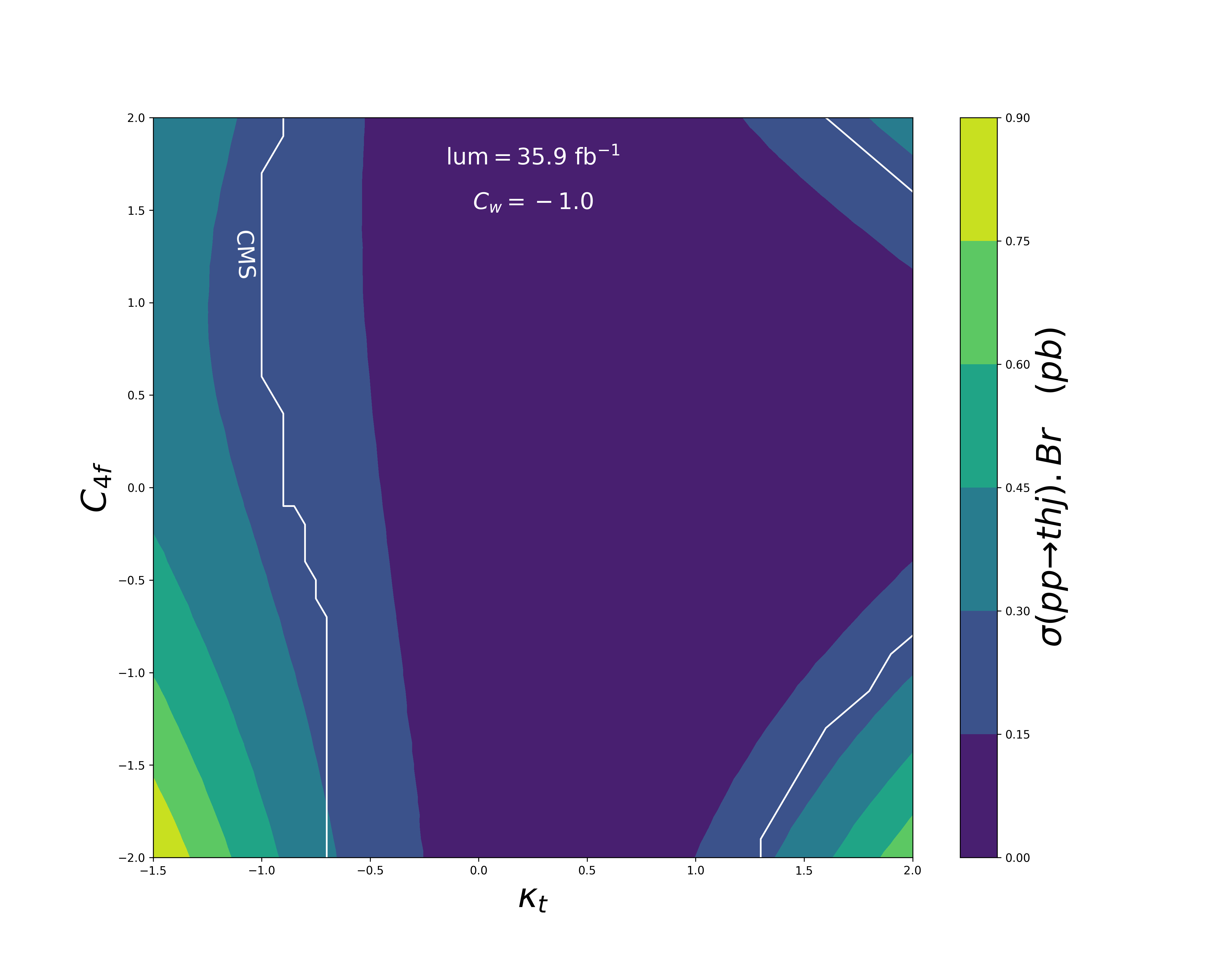}
$$
\caption{Scan for $thj$ production cross-section times branching ratio (in pb) to get SSD signal at LHC in $\wc_{\tt 4f}-\kappa_{\tt t}$ plane for $\wc_{\tt w}=$ 1 (left), -1 (right) at 35.9 fb$^{-1}$ 
luminosity for $\sqrt s=$ 13 TeV. Signal cross-section is shown by colour gradient, where lighter shades 
represent larger cross-section. The bound from observed data in CMS (see Eq.~\ref{bound}) is shown by white lines and the region outside the boundary is excluded by the bound.}
\label{fig:significance1}
\end {figure}

Further, we note the presence of almost uniform $\epsilon_{cut}$ for all the benchmark points, Table 1 suggests that it varies within 0.0014-0.0017.
A constant or slowly varying $\epsilon_{cut}$ basically indicates that the branching ratios and cut sensitivity are almost uniform for different choices of 
EFT benchmark points. Therefore, $\epsilon_{cut}$ can be used effectively to multiply with the {\tt MadGraph} level production cross-section to 
match signal-level SSD events. This methodology is used to put a bound on the cross-section times branching ratio in the EFT 
$\kappa_{\tt t}-\wc_{\tt 4f}$ parameter space as shown in \cref{fig:significance1}. Following the fact that $ \wc_{\tt w}$ affects the cross-section mildly, 
we have kept $\wc_{\tt w}=$ 1 (left), -1 (right); the events are simulated at 35.9 fb$^{-1}$ luminosity at $\sqrt s=$ 13 TeV.
In this figure the cross-section times branching ratio to obtain SSD signal is shown by the colour gradient, where the 
darker shades indicate smaller number of signal events and lighter shades indicate a larger signal cross-section. 
It is easy to appreciate that larger $\kappa_{\tt t},\wc_{\tt 4f}$ provide larger signal cross-section. 
The bound\footnote{There exists almost 13\% uncertainty in the estimation of $N_0$ from Eq.~\eqref{bound}, owing to the mild variation of $\epsilon_{cut}$.} 
obtained from CMS as described in Eq.~\eqref{bound} is shown by the white lines. 
The CMS bound excludes $thj$ cross-section times branching ratio $\gtrsim$ 0.3 pb. This is equivalent to large values of $|\kappa_{\tt t}|$, 
which is mildly dependent on $\wc_{\tt 4f}$ for $\kappa_{\tt t}<0$. The limit $\kappa_{\tt t}<-0.9$ 
is obtained for $\wc_{\tt 4f}=0$, in agreement with the CMS limit. The limit on $\kappa_{\tt t}$ gets tighter for $\wc_{\tt 4f}<0$ and 
relaxed for $\wc_{\tt 4f}>0$, allowing upto $\kappa_{\tt t} \sim -1.0$ (within $-2<\wc_{\tt 4f}<2$). Lastly we note that for $\kappa_{\tt t}>0$, 
the cross-section as well as the CMS limit depends on $\wc_{\tt 4f}$ very crucially. For $\wc_{\tt 4f}=0$, there is no upper bound on $\kappa_{\tt t}$ 
(within the range it has been varied) also in agreement with the CMS analysis. We also see that there is an asymmetry in the dependence on $\wc_{\tt 4f}$ for large 
values of $\wc_{\tt 4f}$ with $\kappa_{\tt t}>0$, which indicates to a non-negligible contribution proportional to $\sim \wc_{\tt 4f}^2$ beyond the interference term 
$\sim \wc_{\tt 4f}\kappa_{\tt t}$, as observed from the selected benchmark points in Table ~\ref{tab:EFT-contribution}. 

We further note that the operators with Wilson coefficient $\wc_{\tt w}$ contributes to $tbW$ vertex and therefore to $t\bar{t}W$ background.
However the contribution of the relevant t-channel graph of $t\bar{t}W$ production is very little, therefore the change is of the order of $0.1$\% of the cross-section 
and can be safely ignored. On the other hand, $\wc_{\tt 4f}$ coupling can contribute to $tb$ production via four fermi interaction (see the Feynman graph in the middle of 
the lower panel of Fig.~\ref{fig:NP.thj}), upon a $W$ radiation, which can contribute non-negligibly $\sim 10$\%
to $t\bar{t}W$ production when no NP is assumed. However, as argued in section \ref{sec:EFT.validity}, the NP that contributes substantially to $thj$ signal 
($Y_1\up{3;1}$, see Eq.~\eqref{mediators} and Eq.~\eqref{s-t}), will contribute to $t\bar{t}W$ via $s$ channel mediation, 
and will be suppressed. On the other hand, $\wc_{\tt 4f}$ contribution to $t\bar{t}Z$ via $Y_1\up{3;1}$  is non-negligible, 
but are suppressed by $b$ parton distribution functions in the initial state. Other NP effects like $Y_1\up{3;2}$ are again s-channel suppressed and produce mild effects 
$\sim 1$\%. Note however, that in estimating above limits, we are talking only about those NP effects which contribute to the `chosen' signal significantly. 
There are dedicated analysis for $t\bar{t}Z, t\bar{t}\gamma, t\bar{t}h$ production in presence of non-negligible EFT contributions, see for example, \cite{CMS:2022hjj}. 
Global fit of SMEFT operators in different channels has been considered in literature exhaustively \cite{Ethier:2021bye}, however a dedicated analysis including 
$tj, thj, t\bar{t}V, t\bar{t}\gamma, t\bar{t}h$ channels is yet to be done, but remains beyond the scope of the present draft.

\section{Upcoming sensitivities at LHC}
\label{sec:reach}

\begin{table}[htb]
\begin{center}
\begin{tabular}{|c|c|}
\hline
\hline
 Signal : $pp \to thj$ & SM background: $pp \to t\bar{t}W$ \cr 
\hline
\hline
 $t\to bW^+\to b\ell^+\nu, ~h\to W^-W^{+*} \to (jj)(\ell^+\nu)$ & $t\bar{t}W\to (bW^-)(bW^+)W \to (b\ell^-\nu)(bjj)(\ell^-\nu)$\cr
\hline
One forward jet & No forward jet \cr
\hline
$b$-jet multiplicity peaks near 1 & $b$-jet multiplicity peaks near 2 \cr
\hline
No reconstructed top in $bjj$ mode & One reconstructed top in $bjj$ mode \cr
\hline
$|\Delta \eta|_{bj_F}$ peaks at large values  & $|\Delta \eta|_{bj_F}$ peaks at smaller values \cr
\hline
$|\Delta \eta|_{\ell j_F}$ peaks at large values  & $|\Delta \eta|_{\ell j_F}$ peaks at smaller values \cr
\hline
\hline
\end{tabular}
\caption{Key distinctive features in SSD events coming from $thj$ signal and $t\bar{t}W$ SM background contributions. The distinctive features 
remain same for $t\bar{t}Z$ also.}
\label{table:sig-back} 
\end{center}
\end{table}

In this section, we refer to the sensitivity with full run-2 data set of LHC at $\sqrt{s}=$ 13 TeV. 
The discovery potential of the signal events depends on the efficiency of reducing the SM background contribution, retaining the signal 
to the extent possible. We observe some key distinctive features between the signal events ($thj$) and the SM background contributions ($t\bar{t}W$) 
as summarized in Table \ref{table:sig-back}. They include: (i) absence of a forward jet in SM ($t\bar{t}W$) background, 
(ii) $b$-jet multiplicity peaks at 1 for $thj$ signal, whereas for background ($t\bar{t}W$) it peaks at 2 having 
two top quarks at the production level, (iii) pseudorapidity difference $\Delta \eta$ between the forward jet and $b$-jet 
($|\Delta \eta|_{bj_F}$) is also expected to be different for signal and background, (iv) $|\Delta \eta|$ between $b$-jet and 
closest lepton ($|\Delta \eta|_{\ell j_F}$) can also provide a distinction between the signal and background. 

\begin{figure}[htb!]
$$
\includegraphics[height=5.5cm]{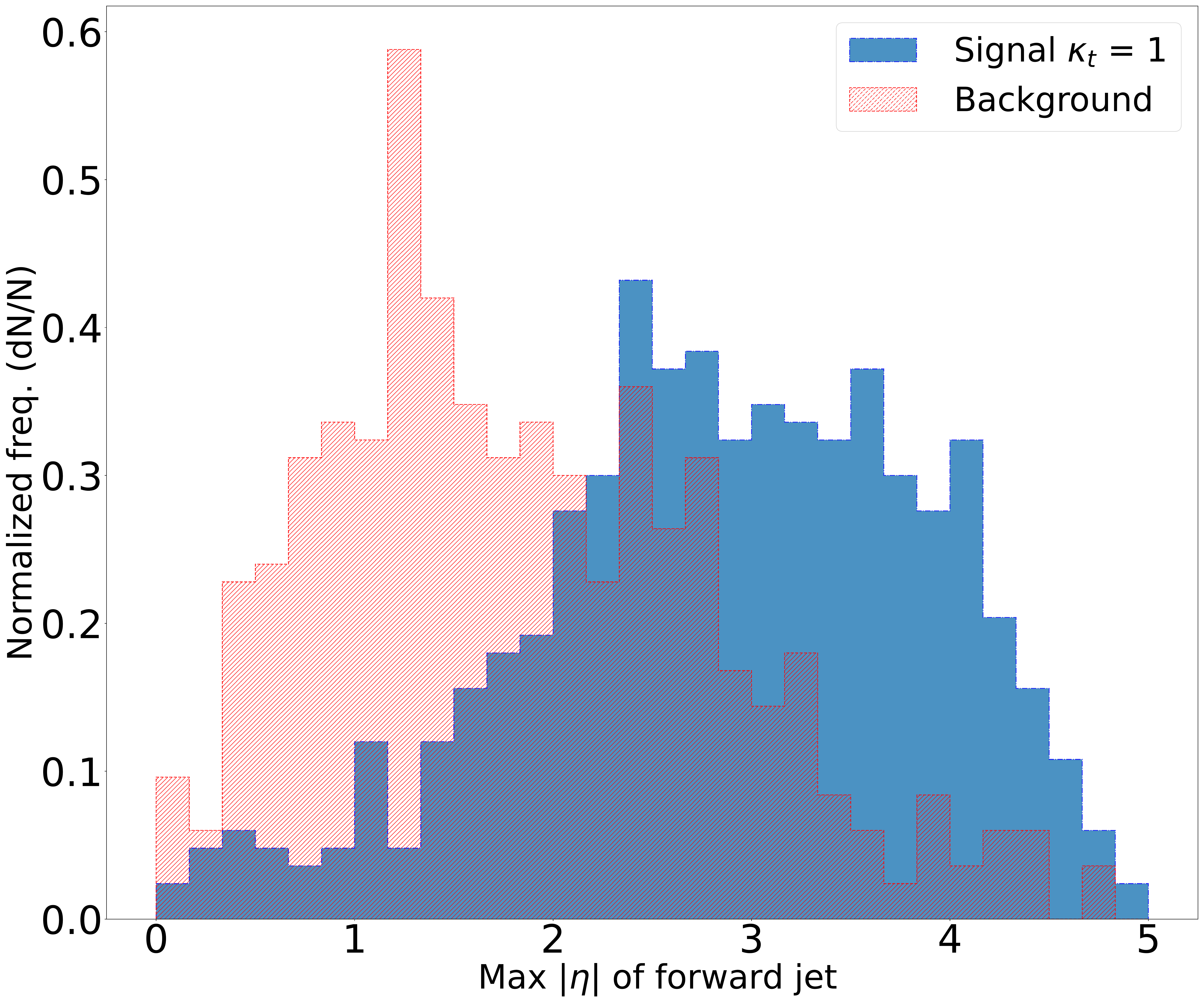}\ \ \ \ 
\includegraphics[height=5.5cm]{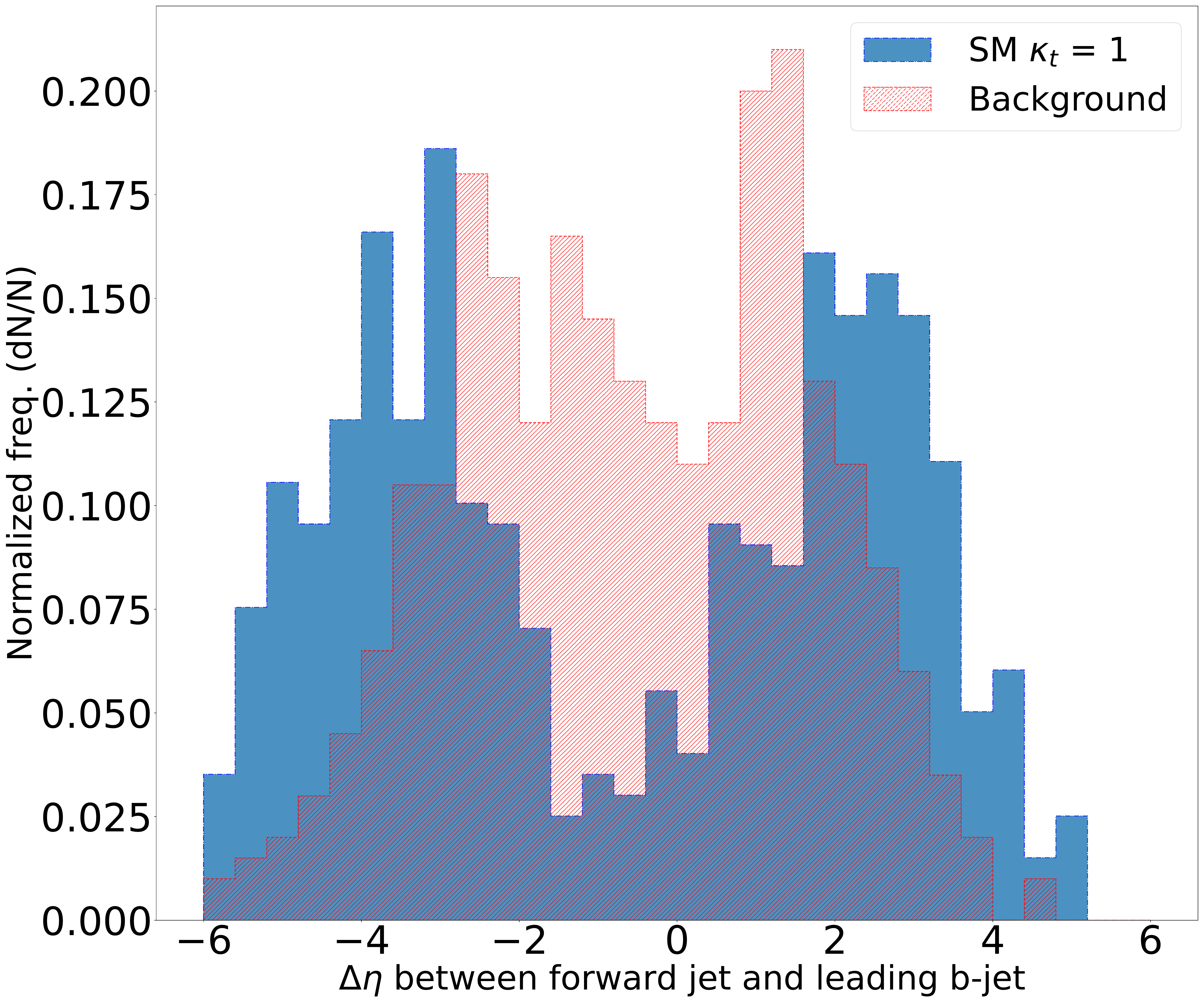}
$$
$$
\includegraphics[height=5.5cm]{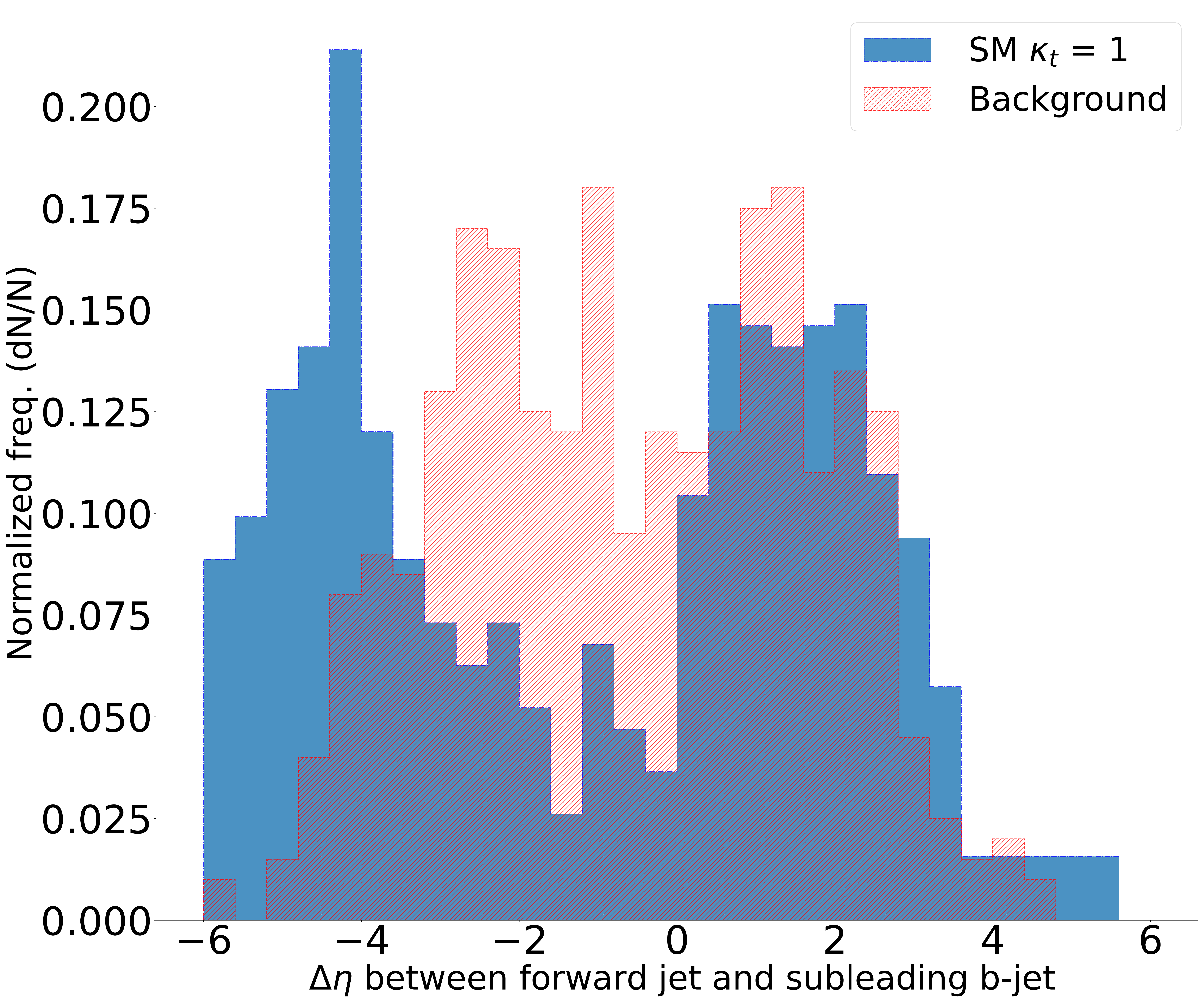} \ \ \ \ 
\includegraphics[height=5.5cm]{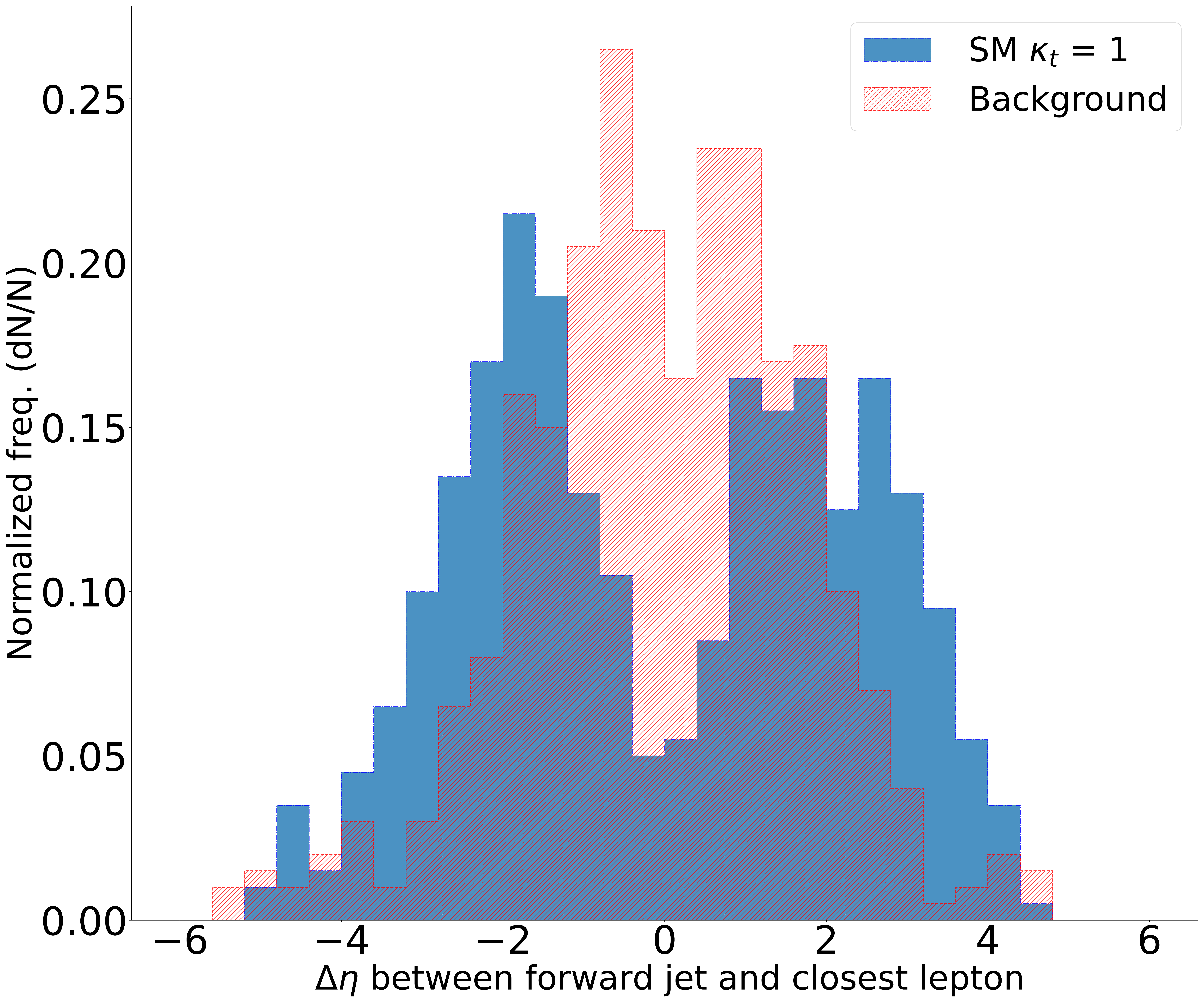}
$$
\caption{Distribution of SSD events coming from SM $thj$ (with $\kappa_{\tt t}=1.0$ in blue shaded region) and SM background ($t\bar{t}W$) in red hatched region 
(all normalized to one) for $|\eta|$ of the forward jet (top left), $(\Delta \eta)_{j_F b}$ with leading $b$ jet (top right), $(\Delta \eta)_{j_F b}$ with sub leading $b$ jet 
(bottom left) and $(\Delta \eta)_{\ell j_F}$ (bottom right) between forward jet and closest lepton at LHC with $\sqrt{s}=$ 13 TeV.}
\label{fig:evt-dist}
\end {figure}

\begin{figure}[htb!]
$$
\includegraphics[height=5.5cm]{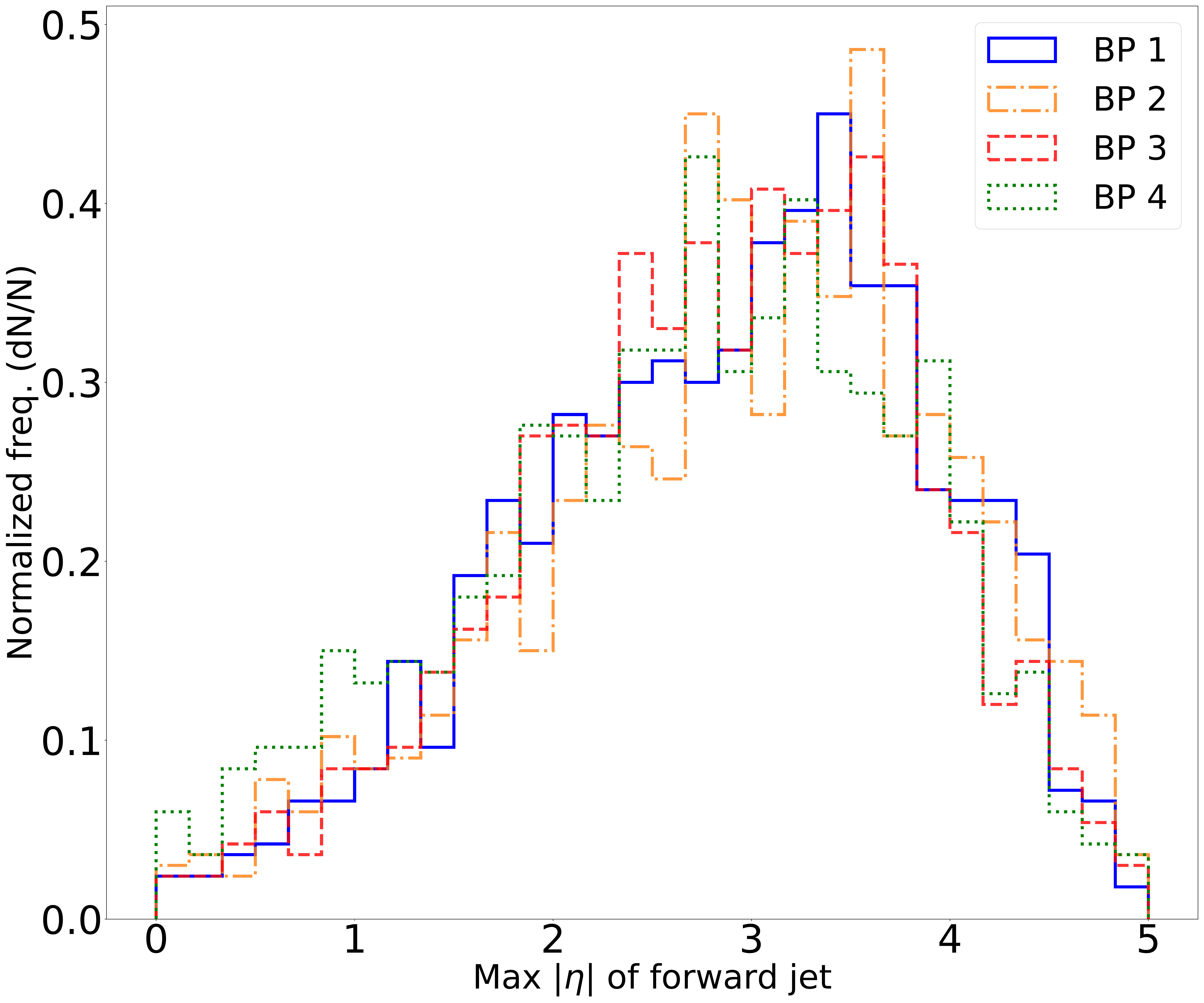} \ \ \
\includegraphics[height=5.5cm]{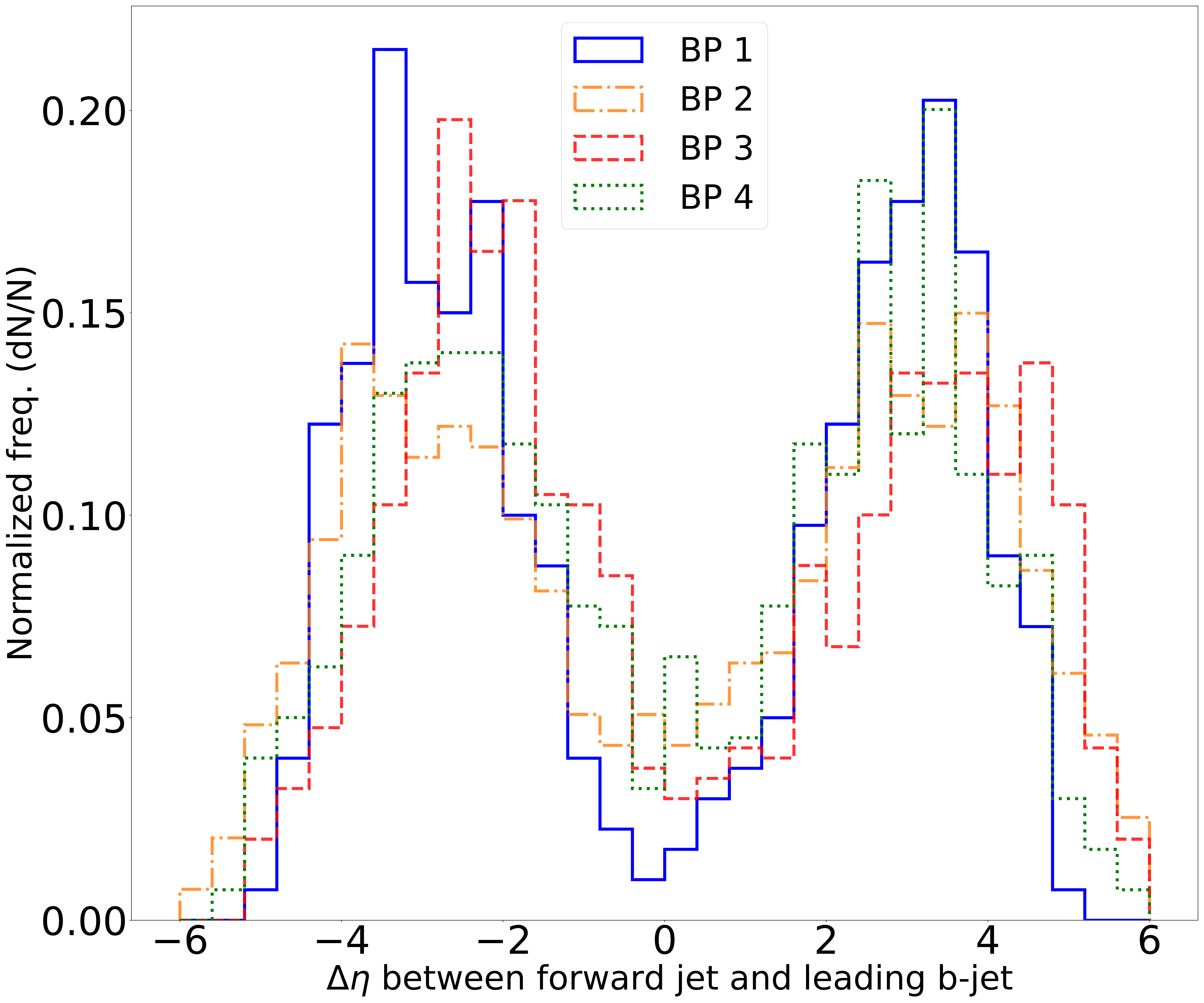}
$$
$$
\includegraphics[height=5.5cm]{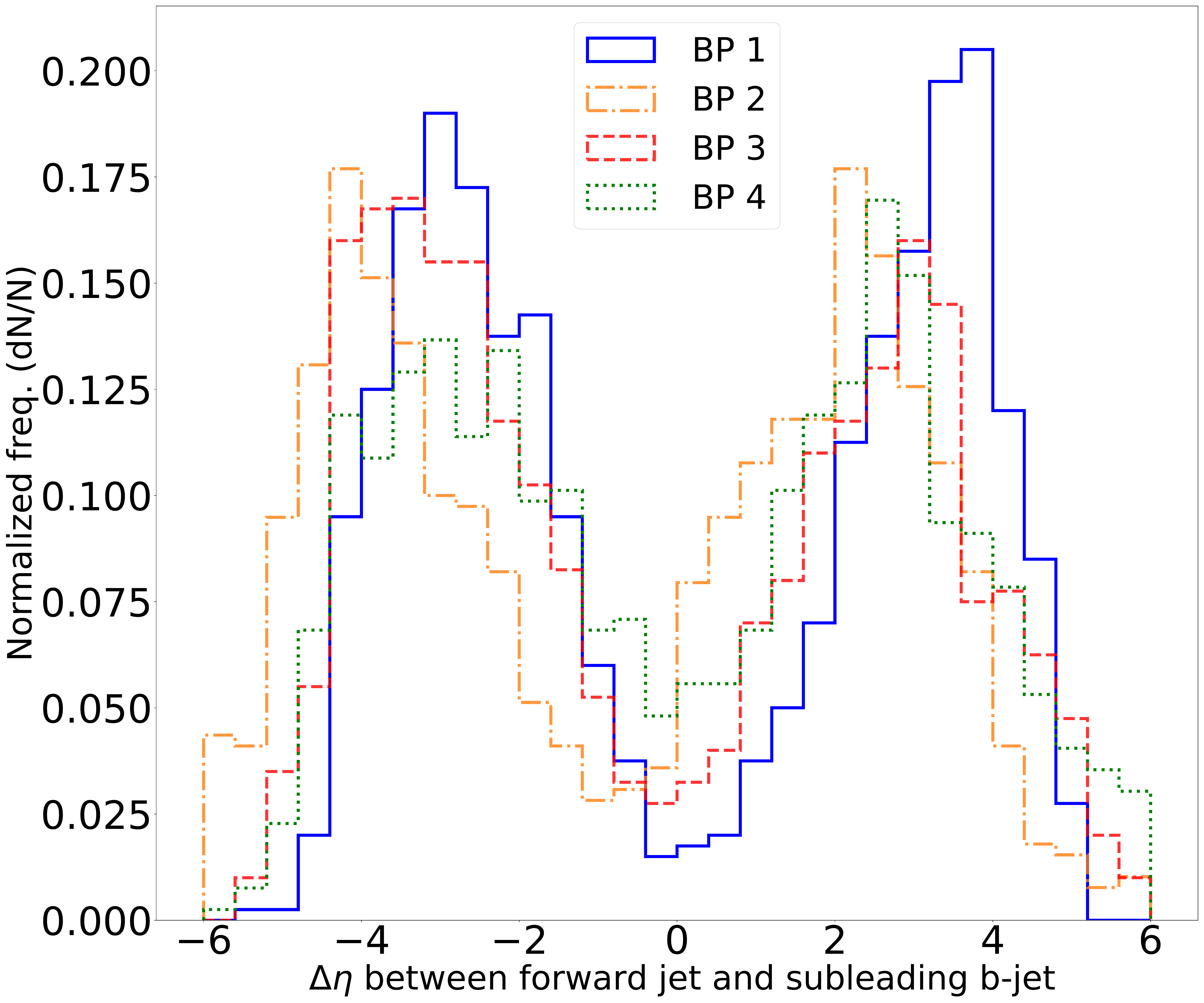} \ \ \ 
\includegraphics[height=5.5cm]{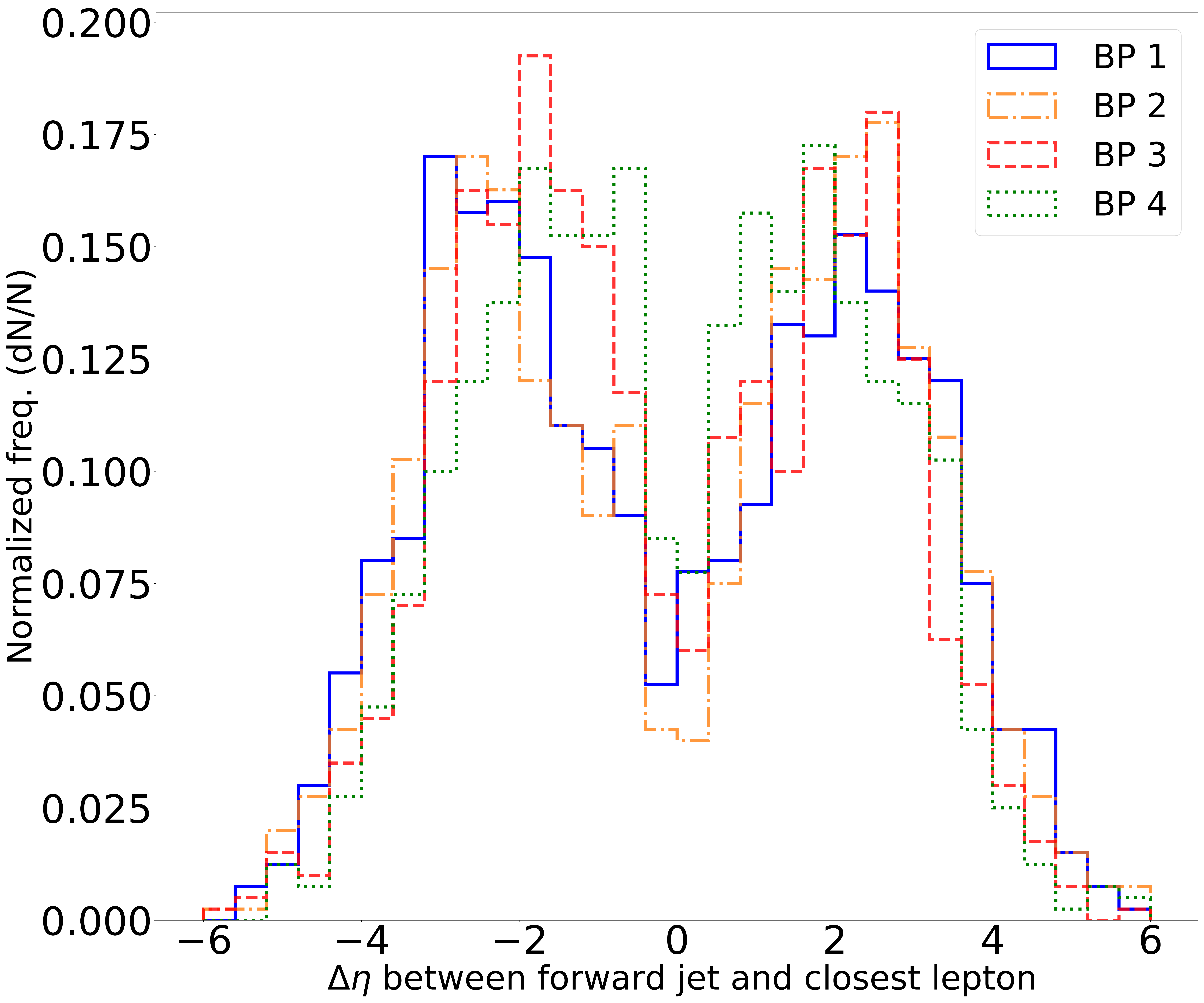}
$$
\caption{Same as Fig.~\ref{fig:evt-dist} for different benchmark points BP1-BP4 without the SM background.}
\label{fig:evt-dist2}
\end {figure}

We show the distributions of SSD events coming from SM $thj$ production with $\kappa_{\tt t}=1.0$ (in blue shaded region) 
and SM background events coming from $t\bar{t}W$ in red hatched regions (all normalized to one) in \cref{fig:evt-dist} for the key variables 
which have been identified to distinguish these two cases; namely, $|\eta|$ of the forward jet (top left), 
$(\Delta \eta)_{j_F b}$ with leading $b$ jet (top right), $(\Delta \eta)_{j_F b}$ with sub leading $b$ jet 
(bottom left) and $\Delta \eta_{\ell j_F}$ between forward jet and closest lepton (bottom right) at LHC with $\sqrt{s}=$ 13 TeV. 
From these distributions, we clearly see that all of these variables can be used judiciously to tame the SM background contribution 
and elucidate the signal events, be it in a cut-based analysis or using machine learning techniques. In \cref{fig:evt-dist2}, we show the 
same distributions for different benchmark points (BP1-BP4) without the SM background. We find that the signal
distributions are not characteristically different for different choices of $\kappa_{\tt t}$ or $\wc_{\tt 4f}, \wc_{\tt w}$. Therefore, uniform signal selection 
criteria can be used for all the allowed regions in the $\kappa_{\tt t}-\wc_{\tt 4f}$ parameter space. Note here that the distributions 
are made with the number of events generated at $\mathcal{L}=35.9$ fb$^{-1}$, but normalized to one.

\subsection{Cut-based analysis}

We chose judicious cuts that distinguish the signal and background events effectively as mentioned 
in Table \ref{tab:hard-cuts}, which can also be appreciated from the distributions shown in \cref{fig:evt-dist}.
The efficiency of these cuts to retain signal and eliminate SM background is shown in Table~\ref{tab:cut-flow} via cut flow for all the 
benchmark points BP0-BP4. $t\bar{t}h$ cross-section at the benchmark points are shown separately in Table \ref{tab:tth}, where 
we have used the SM $K$-factor to normalise the cross-section for all the EFT benchmark points. The maximum $t\bar{t}h$ cross-section 
in the region of parameter space we are interested in is $\sim 567$ fb and the maximum signal contribution after implementing the hard 
cuts is 13 -14 events. Evidently, if this process is included in the background, the estimated signal significance will drop by a few \%. 
For instance, at BP2, the significance reduces to 2.07 from 2.19, whereas for BP3, the significance drops to 0.44 from 0.47 at $\sqrt{s}=$13 TeV 
with 35.9 fb$^{-1}$ luminosity. We also note that the $t\bar{t}$ contribution to the signal can be tamed down after imposing harder set of cuts 
for example, demanding only one b-tagged jet, higher jet rapidity cut like $|\eta|_{j_F}\ge 2.5$, and stronger lepton-jet isolation criteria, 
without much affecting the signal event rate. In the scans for signal significance, we omit $t\bar{t}$ contribution.

We also calculate the signal significance $z(z_1)$ defined by:
\bea
z_1=\frac{S_1}{\sqrt{S+B}},~z=\frac{S}{\sqrt{S+B}},~S_1=S-S^{'}\,;
\label{eq:significance}
\eea  
where $S$ refers to the signal events including EFT, $S^{'}$ refers to signal contribution just with SM and $B$ 
refers to SM background contribution to SSD events. Note here that $z_1$ specifically highlights the effect of EFT 
contribution to the signal, when defined by the difference between signal events produced by EFT 
operators and that from SM as $S_1=S-S^{'}$. A priori, such definition may look unphysical, 
but eventually, all signal events are computed/simulated by some hypothesis and so is this. 

\begin{table} 
\begin{center}
\begin{tabular}{|p{2.5cm}|p{4.5cm}|}
\hline
\hline
Hard Cuts &  $\bullet$ $|\eta_{j_F}|>2.1$ \\
 & $\bullet$ $|\Delta |_{j_F b}>2.0$ \\
 & $\bullet$ $|\Delta |_{j_F \ell}>1.75$.\\
\hline
\hline
\end{tabular}
\end{center}
\caption{Hard cuts used to segregate $thj$ signal and SM background $t \bar{t}W$ for the SSD events.}
\label{tab:hard-cuts}
\end{table}

\begin{table}[htb]
	\begin{center}
\begin{tabular}{|c||c|c|c|c|c|c|c|}
\hline
Cuts & $t\bar{t}W$ & $thj$-SM & BP0 & BP1 & BP2 & BP3 & BP4 \cr 
\hline
\hline
Basic cuts & 199 & 3.9 & 44.6 & 37.7 & 26.2 & 4.7 & 4.0 \cr
\hline
$|\eta_{j_F}|>2.1$ & 63 & 2.8 & 32.9 &27.4 & 19.4 & 3.5 &  3.4 \cr 
\hline
$|\Delta \eta|_{j_F b}>2.0$ & 28 & 2.1 & 25.4 & 20.7 & 14.7 &  3.0 & 2.4 \cr
\hline
$|\Delta \eta|_{j_F\ell}>1.75$ & 18 & 1.3 & 20.8 & 16.9 & 12.0 & 2.1 & 2.3 \cr
\hline
\hline
\end{tabular}
\caption{Number of SSD events at LHC resulting from $thj$ production in SM ($\kappa_{\tt t}=1$), as well as at 
benchmark points BP0-BP4, along with $t\bar{t}W$ background with hard 
cuts applied successively at CM energy $\sqrt s=13$ TeV and $\mathcal{L}=35.9$ fb$^{-1}$.}
\label{tab:cut-flow} 
\end{center}
\end{table}

\begin{table}[htb]
\begin{center}
\begin{tabular}{|c|c|c|c|c|}
\hline
\hline
Benchmarks & \{$\kappa_{\tt t}, \wc_{\tt 4f}, \wc_{\tt w}$\} & $\sigma_{\rm prod}$ (fb) & $N_{\rm basic ~cuts}$ & $N_{\rm final ~cut}$ \cr 
\hline
\hline
SM $t\bar{t}h$ & \{1.0, 0, 0\} &  469.3 & 59.2  &  11.1 \cr
\hline
BP0 &  \{-1.0, 0, 0\} & 469.8   & 59.2 &  11.1 \cr
\hline
BP1 & \{-1.0,1.2,1.0\} &  564.2  &  71.1  & 13.4 \cr
\hline
BP2 & \{-0.5, 1.2,-1.0\}  & 141.8   &  17.9   &  3.4  \cr
\hline
BP3 & \{0.5, 0.4, 1.0\} & 121.1  &  15.3  &  2.9 \cr
\hline
BP4 & \{1.0, 0.8, -1.0\}  & 513.5  & 64.7  &  12.2 \cr
\hline
\hline
\end{tabular}
\caption{$t\bar{t}h$ production cross-section, number of SSD signal events after the basic cuts ($N_{\rm basic ~cuts}$) and hard cuts ($N_{\rm final ~cut}$) 
for SM and EFT benchmark points BP0-BP4 at LHC with $\sqrt s=13$ TeV and integrated luminosity $\mathcal{L}=35.9$ fb$^{-1}$.}
\label{tab:tth} 
\end{center}
\end{table}

\begin{figure}[htb!]
$$
\includegraphics[height=5.5cm]{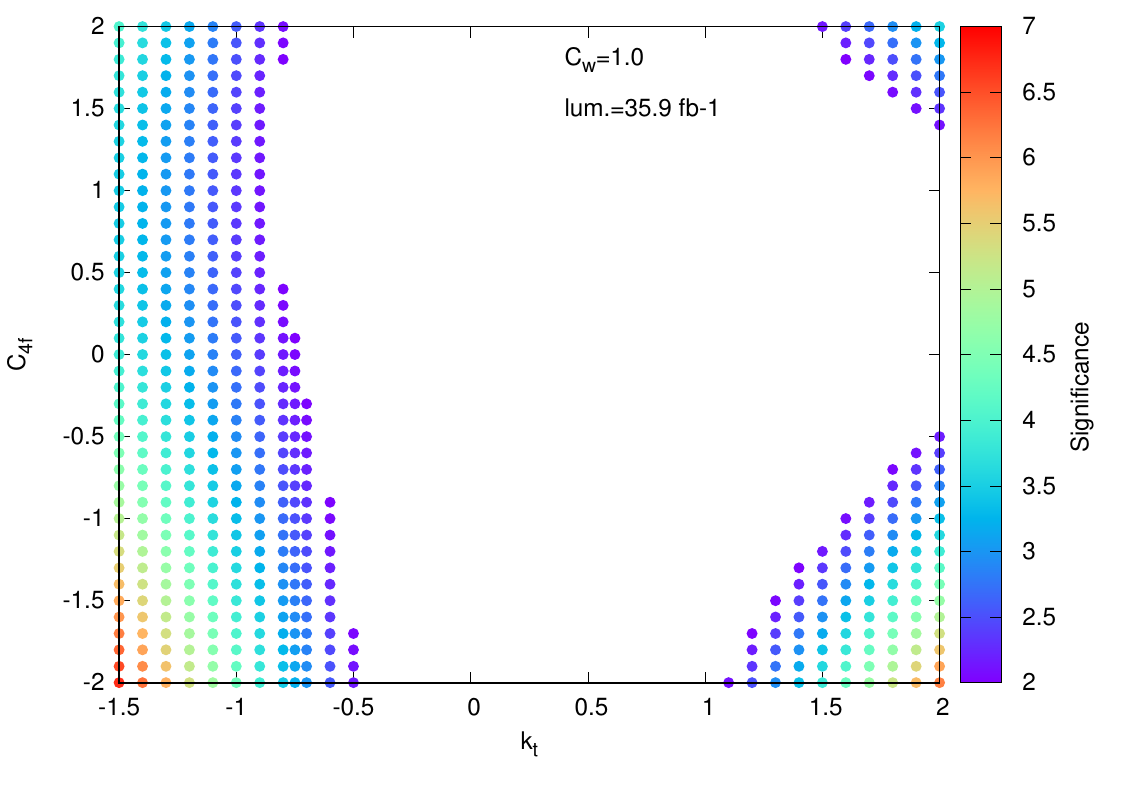}
\includegraphics[height=5.5cm]{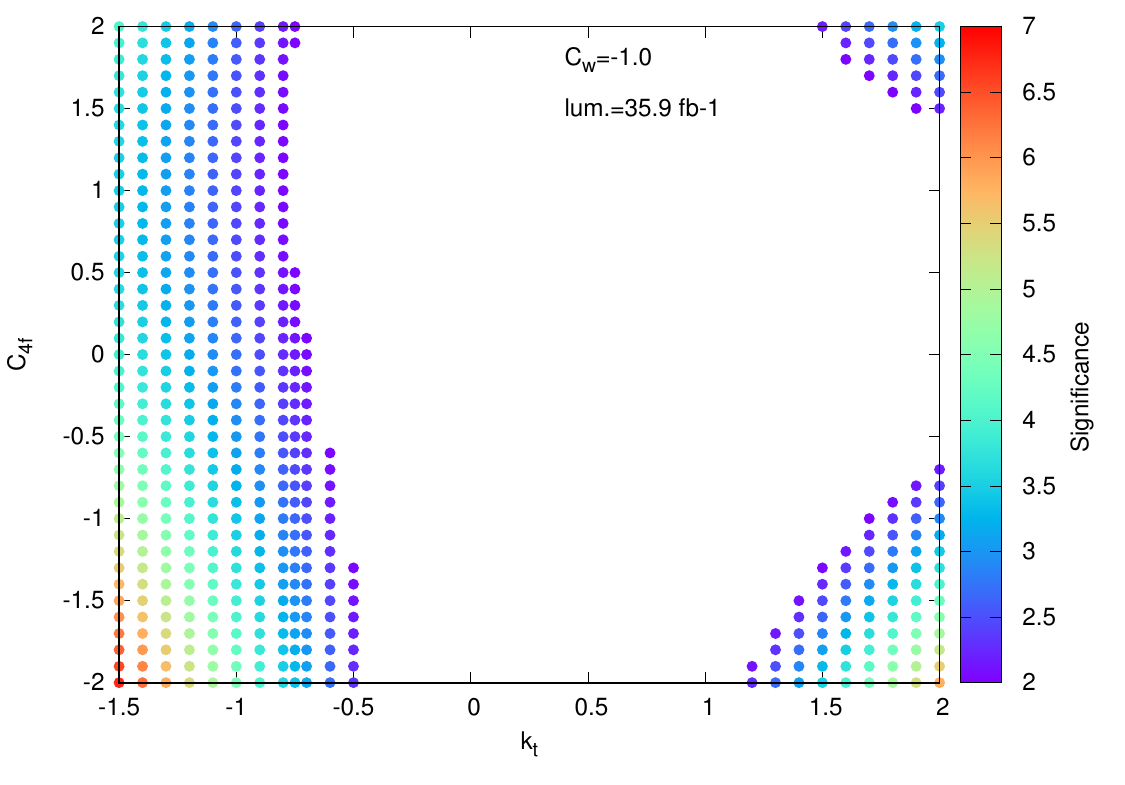}
$$
$$
\includegraphics[height=5.5cm]{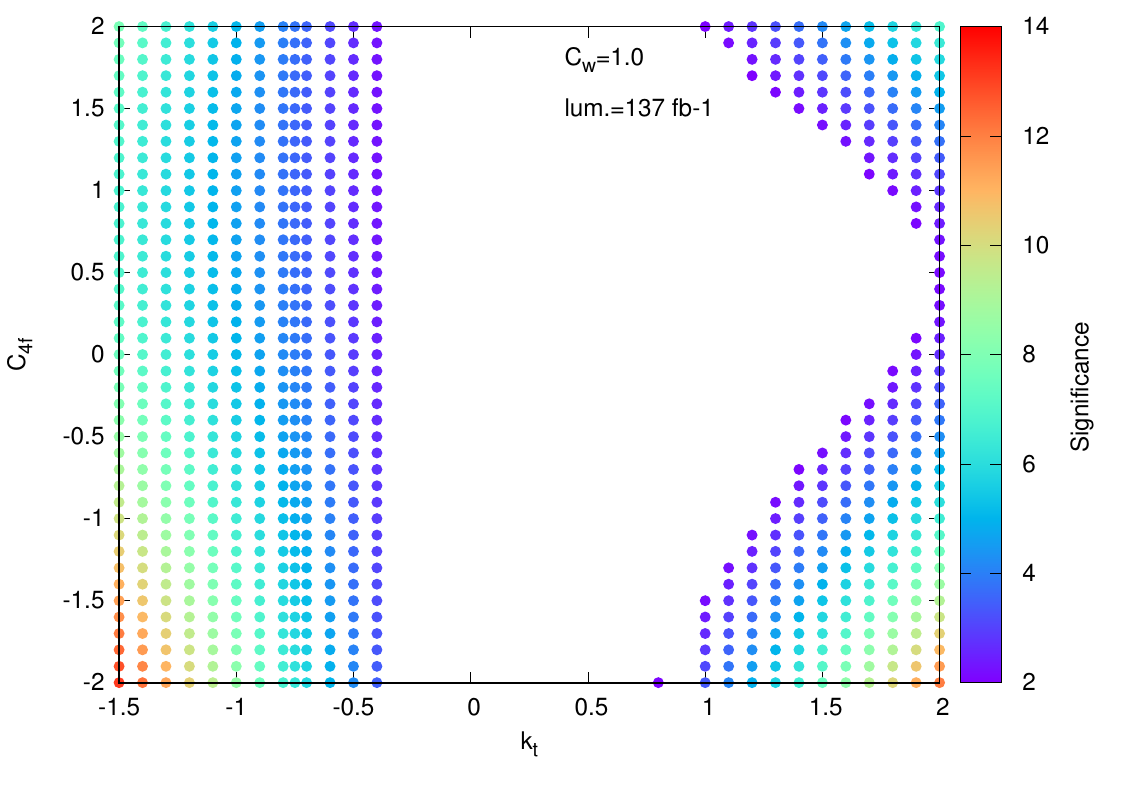}
\includegraphics[height=5.5cm]{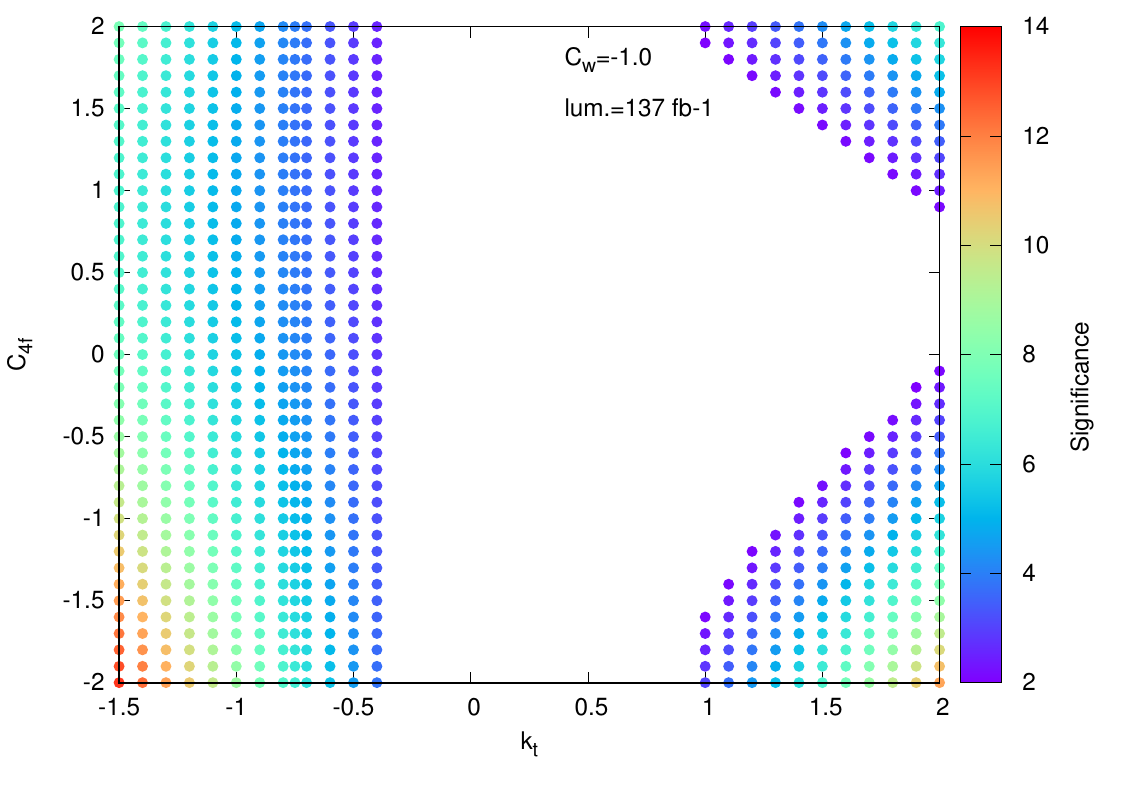}
$$
\caption{2$z_1$ exclusion plot in $\wc_{\tt 4f}-\kappa_{\tt t}$ plane with $z_1$ in colour gradient for constant $\wc_{\tt w}=+1$ (left) and $\wc_{\tt w}=-1$ (right) 
for $\mathcal{L}=$ 35.9 (137) fb$^{-1}$ on top (bottom) panel at $\sqrt s=$ 13 TeV LHC. The color gradient marks regions $\ge 2z_1$.}
\label{fig:significance2}
\end {figure}

We present next in \cref{fig:significance2} the scan for signal significance $z_1$ as colour gradient in $\wc_{\tt 4f}-\kappa_{\tt t}$ plane, for 
$\wc_{\tt w}=$ 1 (left) and $\wc_{\tt w}=$ -1 (right) at $\mathcal{L}=$35.9 (137) fb$^{-1}$ on the top (bottom) panel. The pattern 
remains the same as that of \cref{fig:significance1}, understandably because the signal behaviour remains the same as function of $\wc_{\tt 4f},\kappa_{\tt t}$, while 
the background remain unaffected. The white region falls below 2$z_1$. With larger luminosity, the white region shrinks 
and partially comes under discovery/exclusion limit. For example, the point with $\kappa_{\tt t}=1$ can get excluded in presence of non-zero $|\wc_{\tt 4f}|\sim 1.5$ 
with $\mathcal{L}=$137 fb$^{-1}$. A large portion of $\kappa_{\tt t}<0$ can be excluded or discovered in near future. We should also remember that addition of 
$t\bar{t}h$ contribution to the signal, particularly at the regions of large negative $\kappa_{\tt t}<0$ is significant.

\subsection{Machine learning techniques}
\label{sec:ML}

After estimating a maximally achievable signal significance with a simple cut-based analysis, we further explore the possibility of improving the significance by a 
Machine Learning (ML) technique namely Gradient Boosted Decision Trees (gradient BDT) \cite{chen2016XGBoost} by employing various 
kinematical variables. We use the package \texttt{XGBoost} \cite{chen2016XGBoost} as a toolkit for gradient boosting. We use the same ten observables in the CMS analysis 
\cite{CMS} as input features for the gradient boosting:

\begin{itemize}
\item Number of jets with $p_{T} >  25$ GeV and $|\eta| < 2.4$,
\item Maximum $|\eta|$ of the forward jet,
\item Sum of lepton charges,
\item Number of untagged jets with $|\eta| > 1$,
\item $\Delta \eta$ between forward light jet and leading b-tagged jet,
\item $\Delta \eta$ between forward light jet and subleading b-tagged jet,
\item $\Delta \eta$ between forward light jet and closest lepton,
\item $\Delta \Phi$ of highest-$p_{T}$ same-sign lepton pair,
\item Minimum $\Delta$R between the two leptons,
\item $p_{T}$ of sub-leading lepton.
\end{itemize}
We take an equal number of signal and background events to classify them using the \textit{train} module of \texttt{XGBoost}. For the background, $t\bar{t}W$ and $t\bar{t}Z$ 
events are mixed according to their respective cross-sections. Given the signal distributions are not characteristically different for different choices of $\kappa_{\tt t}$ or $\wc_{\tt 4f}, \wc_{\tt w}$ 
(see Fig. 8), signal events from BP1 are used for training while other benchmark points are used for testing purposes. For training the \texttt{XGBoost} Classifier, we use 10000 
samples each for the signal and background events for BP1 whereas 1000 sample events are used for testing the classifier for all the benchmark points. At first, we use the 
module \texttt{BayesSearchCV} in scikit-optimize \cite{bayessearchcv} library for hyperparameter tuning \cite{claesen2015hyperparameter} to obtain a combination of the 
\texttt{XGBoost} parameters that achieves maximum accuracy to classify the signal and the background events. The module utilises Bayesian Optimization \cite{garnett_bayesoptbook_2022} 
where a predictive model referred to as “surrogate” is used to model the search space and utilized to arrive at good parameter values combination as quickly as possible. The optimized 
parameter values and the corresponding signal efficiencies are listed in tables \ref{table:opt-values} and \ref{tab:bdt-sig} respectively. Signal efficiency (background rejection) for all 
the test benchmark points shows uniformity and averages 75.26 \% (75.48 \%). 

\begin{figure}[htb!]
$$
\includegraphics[height=5.0cm]{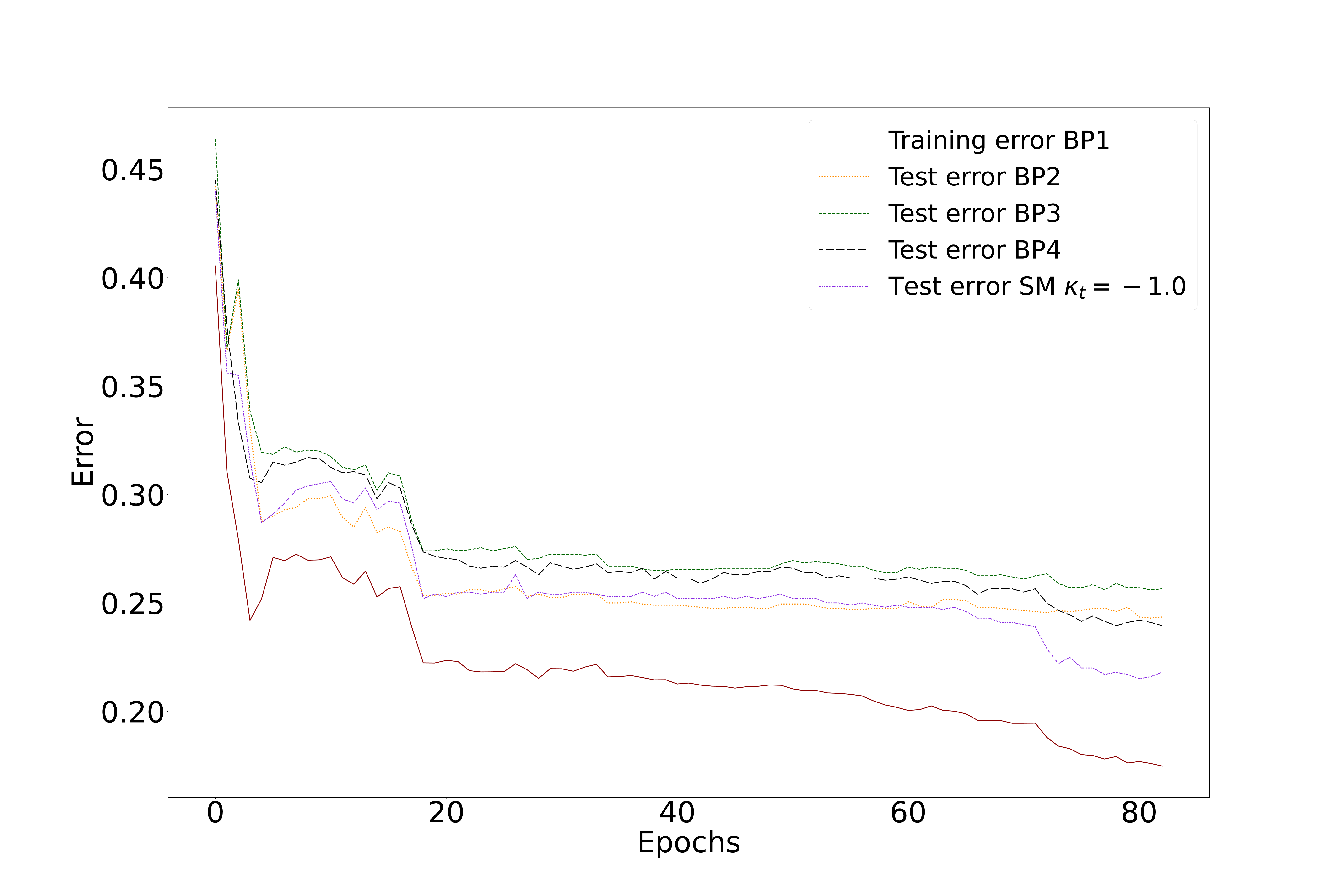}
\includegraphics[height=5.0cm]{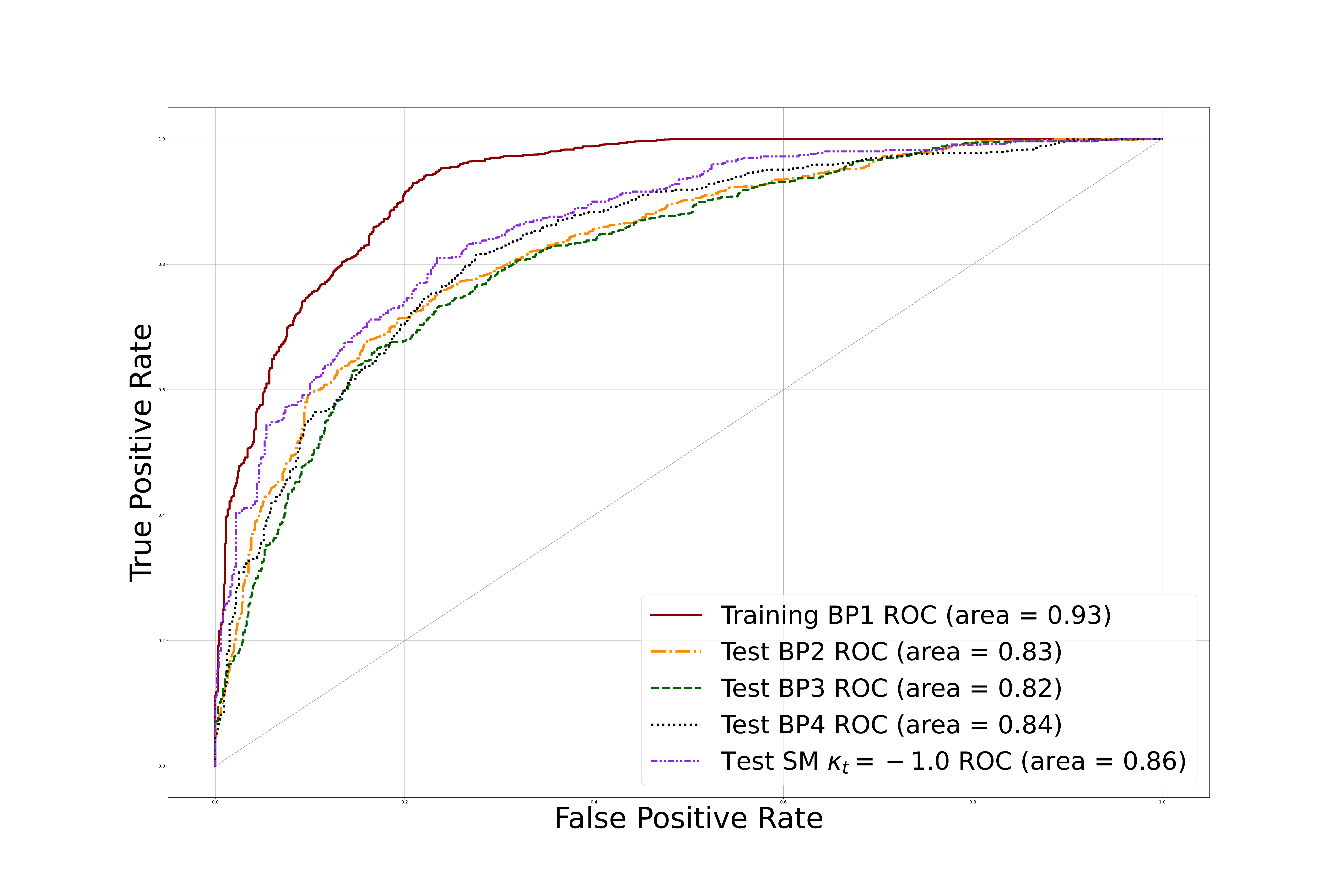}
$$
\caption{Change in loss function with the number of epochs for all the benchmark points (left) and corresponding performance in terms of ROC 
(receiver operating characteristic) for the \texttt{XGBoost}Classifier (right). One epoch is a single run through the entire dataset.}
\label{fig:EFT-cs2}
\end {figure}

In \cref{fig:EFT-cs2}, on the left, we plot the total errors during the training as well as the testing phase as a function of the number of epochs i.e number of runs through the entire dataset whereas on the right we plot the ROC (Receiver Operating Characteristic) curve \cite{10.1016/j.patrec.2005.10.010}. The ROC curve is drawn by plotting the true positive rate (TPR) (signal efficiency) against the false positive rate (FPR) (background rejection) at various threshold settings. The true-positive rate is also known as sensitivity, recall, or probability of detection. The false-positive rate is also known as the probability of false alarm and can be calculated as (1 - specificity). The best possible prediction method would yield a point in the upper left corner or coordinate (0,1) of the ROC space, representing 100\% sensitivity (no false negatives) and 100\% specificity (no false positives). The diagonal line from the bottom left to the top right corners represents random guessing. For the XGBoost classifier, all the test points have similar areas under the ROC curves. Also, the difference in the areas between the training and the test curves is not significant suggesting no overfitting of the model.

We calculate the expected Z-value, which is defined as the number of standard deviations from the background-only hypothesis given a signal yield and background uncertainty, using the {\tt BinomialExpZ} function by {\tt RooFit} \cite{Verkerke:2003ir}.  We use several values for the relative overall background uncertainty, $\sigma_{B}$ = 10\%, 20\%, and 30\% with the currently available integrated luminosity of 35.9 fb $^{-1}$. Clearly, the sensitivity to the NP depends on the relative uncertainty. Keeping that in mind, we analyze all signal channels, assuming that the signal uncertainty is included within $\sigma_{B}$.

\begin{table}[htb]
\begin{center}
\begin{tabular}{|c|c|}
\hline
\hline
 \texttt{XGBoost} Classifier parameters (variable names) & Optimized values \cr 
\hline
\hline
No. of estimators or trees (n$\_$estimators) & 83 \cr
\hline
Learning rate ($\eta$)  & 0.0521 \cr
\hline
 Maximum depth of a tree: (max$\_$depth) & 4.0 \cr
\hline
Subsample ratio of the training instances (subsample) &  0.7565 \cr
\hline
Subsample ratio of columns when constructing each tree (colsample$\_$bytree) & 0.7633 \cr
\hline
subsample ratio of columns for each level (colsample$\_$bylevel) &  0.4436 \cr
\hline
 L2 regularization term on weights ($\lambda$) &  37.0 \cr
\hline
 L1 regularization term on weights ($\alpha$) &  3.0 \cr
\hline
Minimum loss reduction required to make a further partition on a leaf node ($\gamma$)& 3.607 \cr
\hline
Minimum sum of instance weight (hessian) needed in a child (min$\_$child$\_$weight) & 71.0 \cr
\hline
Maximum delta step allowed for each leaf output to be. (max$\_$delta$\_$step) & 4.0\cr
\hline
\end{tabular}
\caption{Optimized parameter values for the \texttt{XGBoost} Classifier tuned using Bayesian Hyperparameter Optimization. The rest of the parameters are set to default options.}
\label{table:opt-values} 
\end{center}
\end{table}

\begin{table}[htb]
	\begin{center}
\begin{tabular}{|c||c|c|c|c|c|c|}
\hline
 & $thj$-SM & BP0 & BP1 & BP2 & BP3 & BP4 \cr 
\hline
N$_{bc}$ & 3.9 & 47.6 & 39.0 & 26.46 & 4.38 & 6.01 \cr
\hline
\multicolumn{7}{ |c| }{\texttt{XGBoost}} \\
\hline
Signal Efficiency & 69.8\% & 81.0\% & 89.5\% & 76.8\%& 74.9\% & 73.8\% \cr

\hline
Background Rejection & 75.4\% & 75.4\% & 80.8\% & 74.5\% & 73.8\% & 78.3\% \cr 
\hline
\hline
\end{tabular}
\caption{Comparison of signal efficiency and background rejection for all the benchmark points and SM-$thj$. The results are obtained using a test dataset of 1000 signal and background events after training the \texttt{XGBoost} classifier on 10000 events from BP1.}
\label{tab:bdt-sig} 
\end{center}
\end{table}

\begin{figure}[htb!]
$$
\includegraphics[height=7.0cm]{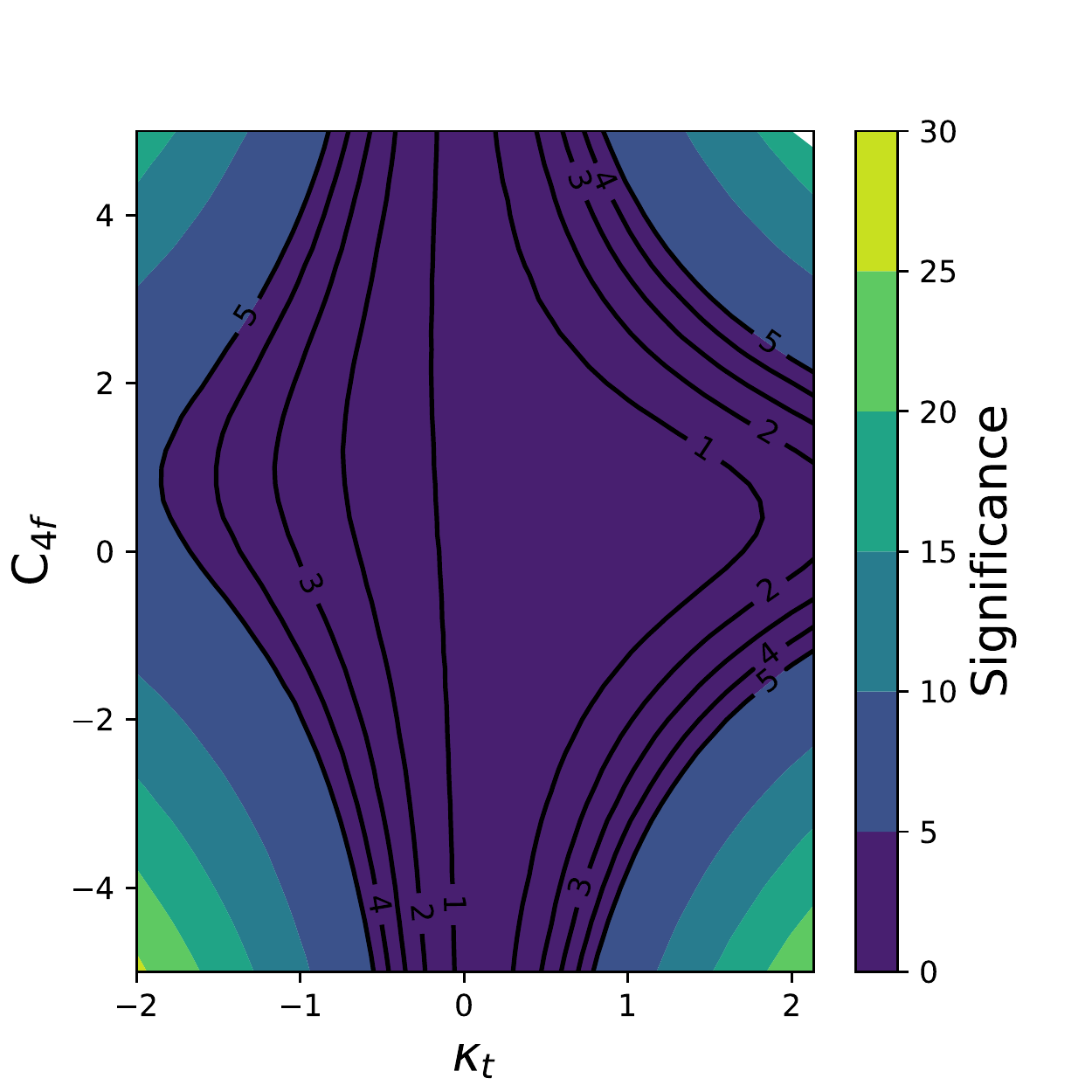}
\includegraphics[height=7.0cm]{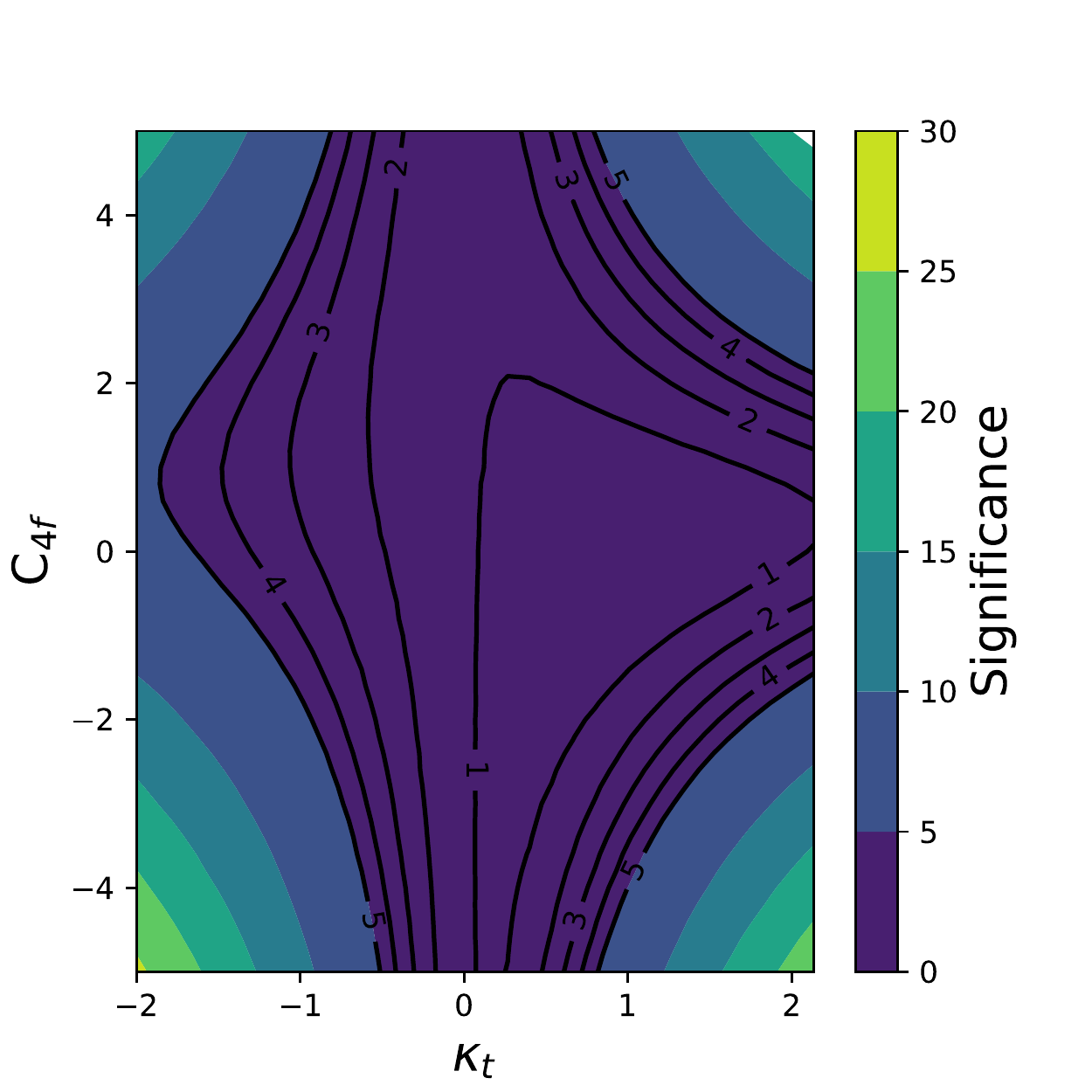}
$$
$$
\includegraphics[height=7.0cm]{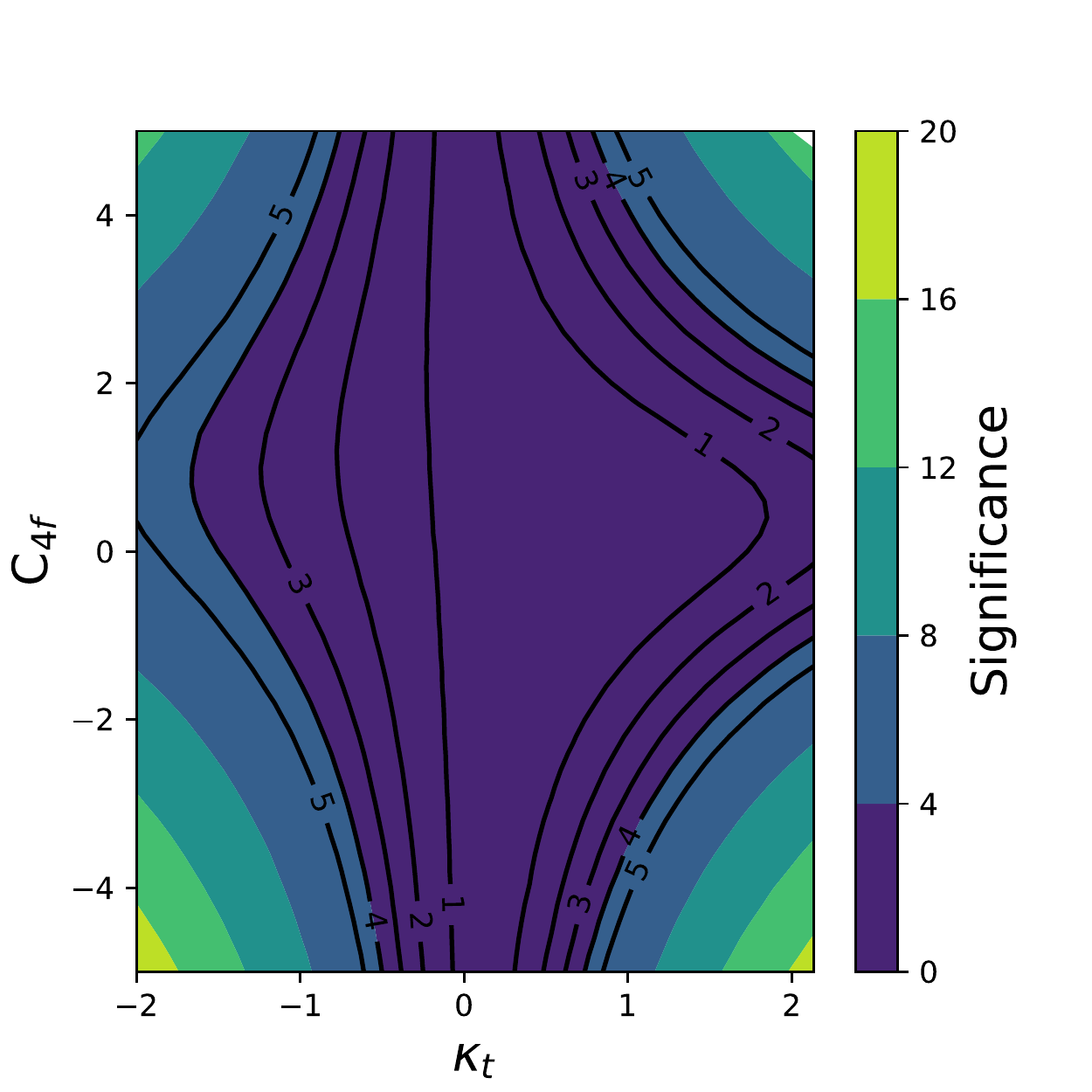}
\includegraphics[height=7.0cm]{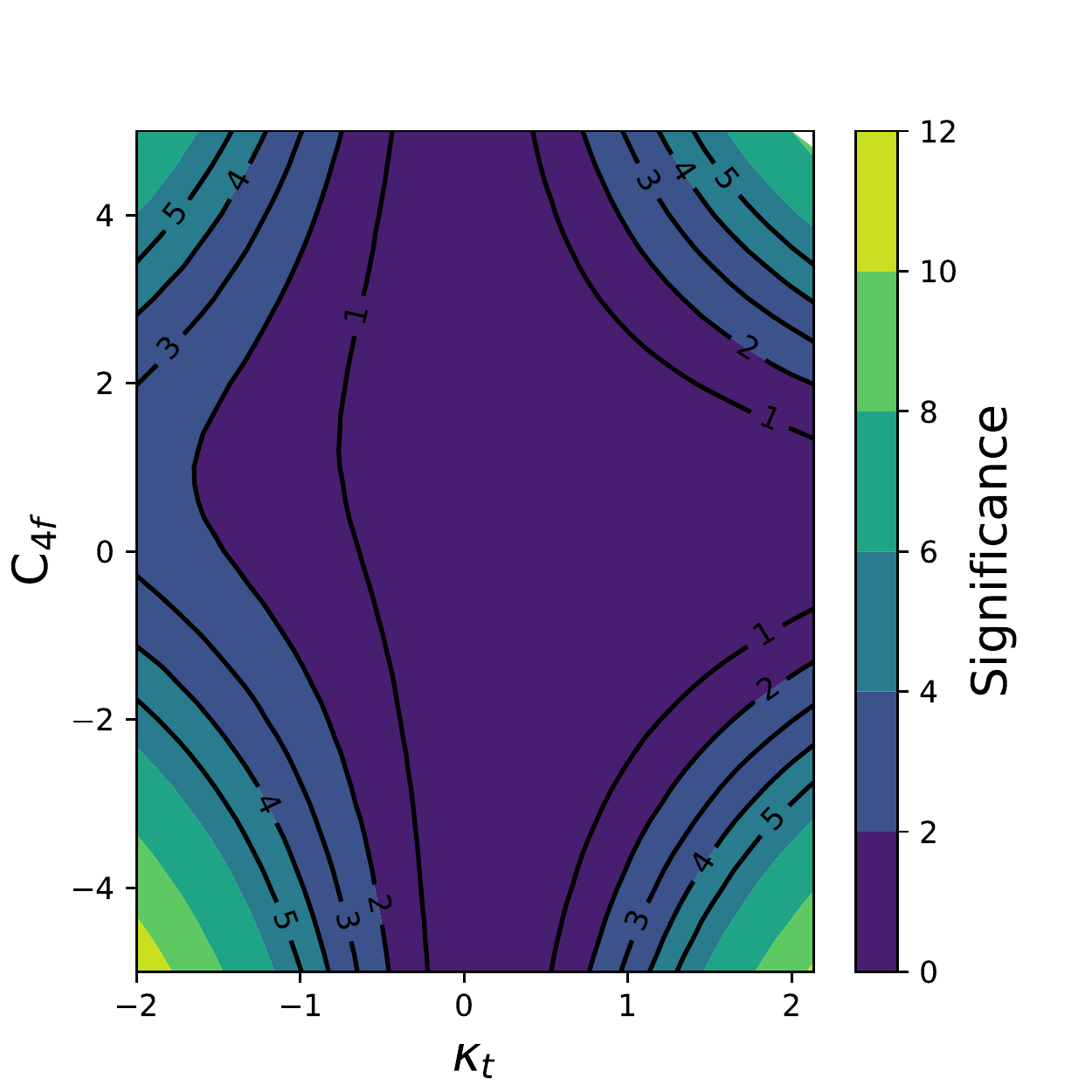}
$$
\caption{Plots with SSD signal significance ($z$) in colour gradient in $\wc_{\tt 4f}-\kappa_{\tt t}$ plane for constant $\wc_{\tt w}=+1$ (left) and $\wc_{\tt w}=-1$ (right) 
constant luminosity 35.9 fb$^{-1}$ at $\sqrt s=$ 13 TeV LHC. Constant significance lines are also shown by black thick lines. 
SM Error has been varied by 10\% in the top panel and 30 \% in the bottom panel.}
\label{fig:significance3}
\end {figure}

Parameter space scan in $\wc_{\tt 4f}-\kappa_{\tt t}$ plane for SSD signal significance ($z$) using ML technique with constant $\wc_{\tt w}=+1$ (left) and $\wc_{\tt w}=-1$ (right) 
with $\mathcal{L}=35.9$ fb $^{-1}$ is shown in \cref{fig:significance3}. Darker shades indicate lower signal significance. 
The pattern is no different than that of \cref{fig:significance1} or \cref{fig:significance2} as the signal dependence on the EFT parameters remains the same. 
We also show constant signal significance ($z$) contours of one sigma, two sigma etc. by black thick lines, which 
show larger and larger concentric curves. In the top panel, the SM background error is varied by 10\% whereas 
in the bottom panel the error is varied in a larger range 30\%. Obviously, this results in smaller significance in lower panel figures, i.e. one sigma 
significance is achieved for larger values of $\wc_{\tt 4f},\kappa_{\tt t}$.

\begin{table}[htb!]
\scriptsize
\hskip-1.0cm\begin{tabular}{|p{0.1\textwidth}|p{0.1\textwidth}|p{0.1\textwidth}|p{0.1\textwidth}|p{0.1\textwidth}|p{0.1\textwidth}|p{0.1\textwidth}|p{0.1\textwidth}|p{0.1\textwidth}|}
\hline 
$\mathcal{L}$ &  \multicolumn{4}{c|}{35.9 fb$^{-1}$}  &  \multicolumn{4}{c|}{140 fb$^{-1}$} \\
\hline
$z$ &  2 & 3 & 4 & 5 & 2 & 3 & 4 & 5 \\
\hline
$\kappa_{t}$ = -1.0 & [0.2, 2.004] & [-1.45, 3.51]  & [-2.44, -] & [-3.25, -] & * & * &  [0.86, 1.38] & [-0.50, 2.62] \\
\hline
$\kappa_{t}$ = 1.0 & [-2.87, 2.96] & [-3.57, 3.64] &  -  &  - & [-1.24, 1.35] &  [-1.70, 1.80]  & [-2.07, 2.16] & [-2.39, 2.46]\\
\hline
$\kappa_{t}$ = 1.5 & [-1.76, 2.17] & [-2.27, 2.62] &  [-2.69, 3.0]  &  [-3.07, 3.33] & [-0.55,1.06] &  [-0.93, 1.40]  &  [-1.22, 1.66] & [-1.47, 1.87]\\
\hline
$\kappa_{t}$ = 2.2 & [-0.88, 1.57] & [-1.30, 1.93] & [-1.65, 2.22] &  [-1.96, 2.46]  & * &  [0.0, 0.80]  &  [-0.37, 1.44] & [-0.63, 1.37]\\
\hline
\end{tabular}
\caption{Exclusion limits on $\wc_{\tt 4f}$  for different values of $\kappa_{\tt t}$ using $\wc_{\tt w}=+1$ using ML techniques at 
$\sqrt{s}=13$ TeV LHC for $\mathcal{L}=35.9,140$ fb$^{-1}$. The $-$ sign indicates the breakdown of the EFT limit whereas $*$ shows no feasible region in the $\kappa_{\tt t}$ - $\wc_{\tt 4f}$ plane.}
\label{exclusion-1}
\end{table}

We then quote the exclusion limits on $\wc_{\tt 4f}$ for constant $\kappa_{\tt t}$ with a constant $\wc_{\tt w}=+1$, 
for different luminosities in Table \ref{exclusion-1} that result from ML-based analysis. 
For example, with $\kappa_{t}$ = -1.0, we see that the $\wc_{\tt 4f}$ can acquire values 0.2 and 2.004 when the signal receives 
2$z$ significance. This is again attributed to both the interference as well as the quadratic contribution of the Wilson coefficient. 
Usually, the points within the quoted maximum limits of $\wc_{\tt 4f}$ provides a lower signal significance with constant $\kappa_{\tt t}$.
The $-$ sign indicates very large values of Wilson coefficients where EFT limit breaks down, and $*$ shows no feasible region in the $\kappa_{\tt t}$ - $\wc_{\tt 4f}$ plane.

\section{Summary and Conclusions}
\label{sec:summary}

In this work, we analyze the single top associated with Higgs production to estimate the limit on top Higgs Yukawa coupling ($\kappa_{\tt t}$) at the LHC in the current and
upcoming sensitivities using run-2 data corresponding to $\sqrt{s}=$ 13 TeV, including SM effective operator contributions upto dimension six.
 All the operators that can contribute to such signal have been considered, taking into account 
limits on such operators from other observables. We see that the most significant contribution comes from the four fermi operators 
$\ocal_{qq}\up1, \ocal_{qq;\,1}\up3,  \ocal_{qq;\,2}\up3$. Apart, the operator $\ocal_{\phi q}\up3$ also contributes, but the effect is much milder. We neglect other 
LG operators as their contributions are suppressed. Together, the Wilson coefficient of four fermi 
operator and the variable Yukawa coupling $\{\wc_{\tt 4f},\kappa_{\tt t}\}$ serve as the major NP parameters of the framework. The other parameter, 
NP scale is kept fixed at $\Lambda=1$ TeV. The effective field theory (EFT) description is validated with $M_{tj}$ invariant 
mass distribution peaking at a much lower value than the chosen NP scale $\Lambda=1$ TeV.

The possible types of heavy physics that generate four fermi operator of the kind $(\bar d_L \gamma_\mu u_L)(\bar t_L \gamma^\mu b_L)$ that contributes to $thj$ 
production at LHC have also been chalked out. Importantly we point out that the NP that contributes potentially to $qb \to q^{'}t$, has a negligible contribution to
$qq^{'} \to t \bar{b}$. Thus the omission of $qq^{'} \to t \bar{b}$ to our signal, in presence of untagged $b$ jet is justified.  

The key variables that distinguish the signal events from the $t\bar{t}W/t\bar{t}Z$ background include $|\eta|$ of the forward jet, 
pseudorapidity difference $(\Delta \eta)_{j_F b}$ between forward jet with both leading and sub-leading $b$ jets, and pseudorapidity 
difference $\Delta \eta_{\ell j_F}$ between lepton and forward jet. The kinematic distributions in presence of the EFT contribution do not significantly differ from 
the corresponding SM ones. An optimised choice of hard cuts on these variables retain signal and provide 5$z$ 
discovery reach for the future luminosities. The same analysis is done with ML technique, which is 
consistent with the cut-based analysis. 

We see that in presence of $\{\wc_{\tt 4f}, \wc_{\tt w}\}$, the 
limit on $\kappa_{\tt t}$ gets stronger for $\wc_{\tt 4f}<0$ and relaxed for $\wc_{\tt 4f}>0$. We also observe an asymmetry in the dependence on $\wc_{\tt 4f}$ for $\kappa_{\tt t}>0$,
which indicates to a potential contribution proportional to $\wc_{\tt 4f}^2$ beyond the interference term with SM. For SM like case with $\kappa_{\tt t}=1$, 3$z$ 
significance can be achieved for both values of $[\wc_{\tt 4f}]=[-2.69, 2.75]$ when $\wc_{\tt w}=+1$, while for $\kappa_{\tt t}=-1$, the same is achieved for much lower 
values of $[\wc_{\tt 4f}]=[0.82, 1.42]$ at integrated luminosity 140 fb$^{-1}$. Thus the limits obtained on $\kappa_{\tt t}$ in absence of EFT operator coefficients like $\wc_{\tt 4f}$ 
are significantly different, making it a necessity to consider them for future analysis of $thj$ process at the LHC.\\

{\bf Acknowledgments}: Subhaditya acknowledges to the Core Research Grant support CRG/2019/004078 from DST-SERB, Govt. of India. 
Sanjoy would like to thank Shivam Verma for various technical helps. We would also like to thank Prof. Kajari Mazumdar and 
Pallabi Das for important clarifications on experimental analysis.

\bibliographystyle{JHEP.bst}
\bibliography{mybib2.bib}

\end{document}